# Suppression and reemergence of directional transport across the ferromagnetic transition in $SrRuO_3$

*Smriti Bhatia*[1], *Neha Sharma*[2], *Srishti Bhardwaj*[2], *Shruti Roy*[3], *Dirk Fuchs*[3], *Tulika Maitra*[2], *Vivek Kumar Malik*[2,*], *Kaushik Sen*,[1,†]

[1]Department of Physics, Indian Institute of Technology Delhi, New Delhi, India

[2]Department of Physics, Indian Institute of Technology Roorkee, Roorkee, India

[3]Karlsruhe Institute of Technology, Institute for Quantum Materials and Technologies, Karlsruhe, Germany

* vivek.malik@ph.iitr.ac.in

† kaushik.sen@physics.iitd.ac.in

Keywords: Anisotropic charge transport, $SrRuO_3$ film, AHE, Positive magnetoresistance

**$SrRuO_3$ is a canonical itinerant ferromagnet that continues to serve as a fertile platform for exploring spin-orbit coupled transport and topological responses in correlated oxide metals. Here, we investigate a (110)-oriented epitaxial $SrRuO_3$ thin film on $DyScO_3$, where crystallographic anisotropy allows for symmetry-resolved charge transport along orthogonal in-plane directions. We uncover two principal findings. First, the resistance above the Curie temperature $T_{c,\mathrm{on}} \approx 165\,\mathrm{K}$ exhibits strong anisotropy, consistent with directionally dependent spin-fluctuation scattering. Below $T_{c,on}$, this anisotropy nearly vanishes, despite persistent structural inequivalence, due to isotropic electron-magnon and electron-electron scattering processes. Second, the application of a magnetic field reveals a striking directional contrast in the linear magnetoresistance $(MR)$: it remains negative and saturates along [001], but changes sign and becomes positive along $[1\bar{1}0]$ below $T^0 = 145\,\mathrm{K}$. In contrast, the quadratic MR term is identical for both directions and peaks near $T^*_{\mathrm{OHE}} = 155\,\mathrm{K}$, where the ordinary Hall coefficient crosses zero, consistent with carrier compensation under a two-band scenario. A sign change in the anomalous Hall signal at $T^*_{\mathrm{AHE}} = 150\,\mathrm{K}$ is consistent with a temperature-dependent redistribution of Berry-curvature contributions near the Fermi level. These results establish ferromagnetic ordering as the primary driver of the observed directional magnetotransport in $SrRuO_3$.**

## Introduction

$SrRuO_3$ (SRO) is a canonical $4d$ itinerant ferromagnet that offers a fertile platform to investigate the interplay of magnetic order, electronic correlations, and spin-orbit coupling (SOC) in correlated oxide metals [1]. With a moderately high Curie temperature near 160 K,

SRO exhibits a wide range of exotic transport and topological phenomena. These include a prominent anomalous Hall effect (AHE) linked to Berry curvature near the Fermi level [2–4], unsaturated linear positive magnetoresistance identified as a transport signature of Weyl-derived electronic states [5], and skyrmion-like spin textures in thin films [6] and heterostructures [7]. The ability to control structure and symmetry through epitaxial strain has further expanded its potential as a model system to probe symmetry-sensitive quantum transport [3].

In the paramagnetic phase, SRO enters a 'bad metal' regime in which the resistivity rises nearly linearly with temperature and exceeds the Mott-Ioffe-Regel limit without saturation [1],[8],[9]. This transport behavior is widely interpreted in terms of strong spin fluctuations, consistent with Moriya's theory for itinerant magnets [10]. Our measurements strongly support this scenario. Above the ferromagnetic ordering temperature $T_{c,\mathrm{on}}$, the resistance $R(T)$ shows marked anisotropy between the in-plane [001] and $[1\bar{1}0]$ directions, signaling that spin-fluctuation scattering is not only active but also directionally dependent, likely reflecting orbital anisotropy, Fermi velocity mismatch, and SOC-driven entanglement [11–13]. Fang et al. reported resistivity and Hall measurements on both single-crystalline SRO and SRO thin films and found qualitatively similar features in the two cases [2]. Despite these insights, the influence of structural anisotropy on transport across the ferromagnetic transition has not been systematically studied. In particular, it remains unclear how long-range magnetic order modifies the directional dependence of scattering, and whether Berry-phase effects play a role in such symmetry-resolved transport. Oriented thin films of SRO provide a unique opportunity to address these questions by enabling access to crystallographically inequivalent directions within a single-domain sample.

In this study, we investigate a (110)-oriented epitaxial SRO thin film on $DyScO_3$ and examine the evolution of resistance and magnetoresistance $(MR)$along the in-plane $[001]$ and $[1\bar{1}0]$ directions. These directions are structurally inequivalent due to the orthorhombic crystal symmetry of SRO on $DyScO_3$, allowing us to probe symmetry-resolved transport with high precision. Two key findings emerge. First, the resistance $R(T)$ exhibits clear anisotropy above $T_{c,\mathrm{on}}$, which becomes strongly suppressed below the transition. We concluded that this is an indication of a crossover from directionally dependent spin-fluctuation scattering to isotropic contributions from electron-magnon and electron-electron interactions. Second, although the zero-field $R(T)$curves appear nearly isotropic below $T_{c,\mathrm{on}}$, application of a magnetic field reveals pronounced anisotropy in the MR: the linear MR remains negative along [001] but becomes positive along $[1\bar{1}0]$ below $145\,\mathrm{K}$ (labeled as $T^0$). In contrast, the quadratic component of MR is identical in both directions and peaks at the same temperature where the ordinary Hall coefficient vanishes.

To understand the origin of these $MR$ trends, we analyzed the Hall resistance as a function of temperature. The ordinary Hall coefficient $R_o$ changes sign at $T^*_{\mathrm{OHE}} = 155$ K, indicating a transition from hole-dominated to electron-dominated Hall response. Slightly below this, the anomalous Hall component $R_sM$ also reverses sign at $T^*_{\mathrm{AHE}} = 150$ K, consistent with a Berry curvature redistribution near the Fermi level. These observations establish that the $MR$ anisotropy develops in the same temperature window as the electronic reconstruction driven by exchange splitting. The simultaneous reversal of magnetoresistance and anomalous Hall response indicates that the change in MR sign originates from a reconstruction of the Berry curvature near the Fermi level, which leads to direction-dependent charge transport.

Together, these results show that the ferromagnetic transition in SRO activates a narrow temperature window where spin fluctuations, band topology, and crystallographic anisotropy interact to shape magnetotransport. By systematically disentangling these effects in a single epitaxial film, our study provides new insight into the symmetry-resolved dynamics of correlated oxide metals and highlights the central role of magnetic order in modulating topological transport.

**Results**

Figure 1(a) schematically illustrates the crystallographic arrangement of a 50 nm thick (110)-oriented $SrRuO_3$ (SRO) film grown on a (110)-oriented $DyScO_3$ (DSO) substrate. Both SRO and DSO crystallize in an orthorhombic perovskite structure with space group *Pbnm*, corresponding to the crystallographic point group $D_{2h}$. The crystallographic long $c$-axis lies in-plane along [001], while the $a$- and $b$-axes are tilted 45° with respect to the film normal [110] (detailed in sec. 1.2 in supplementary information). Thus, this epitaxy defines two orthogonal in-plane transport directions, [001] and $[1\bar{1}0]$, that are crystallographically inequivalent. While the lattice symmetry of SRO remains unchanged, its ferromagnetic ordering breaks time-reversal symmetry, leading to a magnetic symmetry set by the magnetization direction.

Figure 1(b) shows zero-field resistance ($R$) as a function of temperature ($T$) along orthogonal in-plane directions [001](red circles) and $[1\bar{1}0]$ (blue circles). In both directions, $R$ increases non-monotonically with temperature, consistent with the metallic character of SRO. The residual resistivity ratios (RRR) are 5.2 for [001] and 4.7 for $[1\bar{1}0]$, falling within the typical range of 3-20 for epitaxial films [14–16]. Exceptionally, an MBE-grown SRO film on DSO has been reported with a much higher RRR of ~80.1 [17]. A pronounced drop in $R$ occurs below the ferromagnetic (FM) transition temperature $T_{c,on} \approx 165$ K, identified by the onset of the peak in $dR/dT$ (inset), marking the temperature where spin-disorder scattering first begins to collapse [1]. Above $T_{c,on}$, $R(T)$ is nearly linear and strongly direction-dependent, reflecting distinct scattering environments along the two orthogonal in-plane directions. Below $T_{c,on}$, this anisotropy vanishes and $R(T)$ becomes nearly identical along both directions, despite the persistent crystallographic inequivalence, demonstrating that FM order strongly influences charge transport in the ferromagnetic state. Similar anisotropy has been observed in other monodomain (110)-oriented SRO thin films, where the in-plane $R(T)$ anisotropy decreases progressively upon cooling and becomes significantly suppressed below $T_{c,\mathrm{on}}$ [8],[18] (shown in Figure S5 sec. 3 SI).

To clarify the low-temperature behaviour we inspected the derivative $dR/dT$. Within the ferromagnetic state, two distinct scattering regimes are evident in agreement with existing reports [9],[19] : a first interval from 4 to $\sim$ 40 K and a second from about $\sim$60 to $\sim$140 K, with a crossover near 50 K. This 50 K anomaly matches the Fermi liquid to non-Fermi liquid (FL to nFL) crossover previously reported [1],[20]. In the 4-40 K interval, the resistance follows a purely quadratic $T$ dependence ($R \propto T^2$), consistent with typical Fermi-liquid behaviour governed by electron-electron scattering [21]. It is to be noted that momentum-conserving coherent electron-magnon scattering can also yield $R \propto T^2$ [20]. Between 60 K and 140 K, the data are best described by a $T^{3/2}$ power law (nFL regime), which is consistent with reported results [20]. Mills *et al.* showed that incoherent electron-magnon scattering, which does not conserve momentum, leads to such $T^{3/2}$ dependence of $R(T)$ [22].

Above $T_{c,on}$, SRO enters a paramagnetic bad-metal regime in which the resistivity increases almost linearly with temperature (empirically $T^{0.7} - T^{0.8}$) and exceeds the Mott-Ioffe-Regel limit without saturation [1],[8],[9]. Klein *et al.* experimentally established that spin fluctuations and magnetic ordering strongly affect electrical transport in SRO, particularly near $T_{c,on}$ [8]. Consistently, Mazin and Singh theoretically showed that $R(T)$ in the paramagnetic phase is strongly influenced by electron-paramagnon scattering [23]. Our data are consistent with this picture, as the near-linear resistance above $T_{c,on}$ is compatible with persistent spin-fluctuation scattering in the paramagnetic phase.

Figure 2(a) and (b) show isothermal magnetoresistance $\left(MR = \frac{R(\mu_0 H)-R(0)}{R(0)} \times 100\ \%\right)$ for currents along [001] and $[1\bar{1}0]$, respectively. The magnetic field, $\mu_0 H$ is applied perpendicular to the film plane, i.e. along [110]. For $T > T_{c,on}$, the $MR$ is extremely weak. A sizeable negative $MR$ ($nMR$) appears at temperatures below $T_{c,on}$; its magnitude is maximal near the transition, reflecting the field-induced suppression of spin fluctuations [24]. On further cooling down to 100 K, $nMR$ evolves smoothly from a non-linear field dependence into a strictly linear one. Along [001], the $MR$ remains negative down to the lowest measurement temperature of 100 K, whereas along $[1\bar{1}0]$ it passes through zero and becomes positive below $T^0 \approx 145$ K. At the sign-reversal temperature, the $MR$ amplitude is as small as in the high-temperature paramagnetic phase.

The observed $MR$ vs. $\mu_0 H$ curves are well described by a two component model with linear (even function) and quadratic terms in $\mu_0 H$, as shown in eq. (1) (see Figure 2(a) and (b)).

$$\mathrm{MR}(\mu_0 H) = A(T)|\mu_0 \mathrm{H}| + C(T)(\mu_0 H)^2 \quad (1)$$

The quadratic coefficient $C(T)$ is due to applied-field induced Lorentz force on the electron- and hole-like charge carriers, typical for semimetal or doped semiconductors where both types of charge carriers exist [25]. In SI (sec. 2), we have given the expression for of $C(T)$ as a function of charge densities. The linear coefficient $A(T)$ in ferromagnets can arise from multiple mechanisms. The two primary contributions are: (i) the suppression of spin-disorder scattering under magnetic field, particularly due to reduced electron-magnon spin-flip scattering below $T_{c,on}$, as established in conventional $3d$ ferromagnets [26] and (ii) the presence of topologically nontrivial band structures, which give rise to finite Berry curvature and an anomalous velocity contribution to magnetotransport [27]. The latter is inherently momentum-dependent and can be highly anisotropic, since the distribution of Berry curvature in the Brillouin zone is strongly shaped by crystal symmetry, spin-orbit coupling, and the orientation of the applied magnetic field [27].

Figure 2(c) and (d) show the temperature evolution of $A(T)$ and $C(T)$, respectively. Above $T_{c,on}$, $A(T)$ is almost zero. Immediately below $T_{c,on}$, it becomes large and negative, reaching its most negative value near the FM transition, precisely at $T^{A_{\min}} \approx 155$ K. Although the initial drop in $A(T)$ is common to both the directions, their low-temperature behaviours diverge. Along [001], $A(T)$ saturates at a constant negative value; whereas along $[1\bar{1}0]$, $A(T)$ increases continuously, crosses through zero at $T^0$ =145 K, and remains positive with no clear saturation within the measured temperature window down to 100 K; the trend suggests that any eventual plateau along $[1\bar{1}0]$ would occur only at lower temperatures.

By contrast, $C(T)$ is remarkably identical for the two in-plane directions. It peaks sharply and positively at $T^{C_{\max}} \approx 155$ K, slightly below our $T_{c,on}$, then decays toward low temperatures, eventually attaining a small negative value, where $MR$ is dominated by the linear coefficient. Thus, the strikingly different temperature dependences of $A(T)$ and the identical behaviour of $C(T)$ highlight the role of the underlying crystallographic anisotropy: the magnetic field re-introduces the directional inequivalence that appeared to vanish in zero-field transport, and does so solely through the linear component of $MR$, which is linked either to electron-magnon spin flip scattering or to Berry-curvature-induced anomalous carrier velocities.

To clarify the connection between the observed magnetoresistance and underlying charge carriers we measured the Hall resistance at the same temperatures. Figure 3(a) and (b) plot $R_{\mathrm{Hall}}$ versus $\mu_0 H$ applied along surface normal [110]. Well above $T_{c,on}$, the curves are purely ordinary: straight lines with positive slope that identify hole-dominated charge transport. When the temperature is lowered through $T_{c,on}$, a square contribution develops in $R_{Hall}$ vs. $\mu_0 H$ around zero field, producing a hysteresis loop that closes once the field is large enough (only ~0.15 T at 100 K) to saturate the magnetization ($M$). Beyond this closing point the response returns to a single linear trend with a finite slope due to the ordinary Hall effect.

The high field portions, where $M$ is saturated (closed hysteresis loop), were fitted using eq. (2).

$$R_{\mathrm{Hall}}(\mu_0 H, T) = R_o(T)\mu_0 H + R_s(T) M(\mu_0 H, T) \quad (2)$$

Here $R_o(T)$ is the ordinary Hall coefficient and $R_s(T)M$ represents the anomalous Hall part. Figure 3(c) shows $R_o(T)$ increasing sharply as the system approaches $T_{c,on}$ from above, reaching a peak near 160 K. It then drops rapidly, crossing zero at $T^*_{OHE} \approx$ 155 K, and becoming negative at lower temperature. $T^*_{OHE}$ is the characteristic temperature where compensation of electron and hole contributes to the Hall effect. In the semiclassical two-band model, $C(T)$ is maximized at $T^*_{OHE}$ where the carrier conductivities balance ($n_e \mu_e \approx n_h \mu_h$), as detailed in sec. 2 in SI [25].

Since the exact value of $M$ is not known, Figure 3(d) shows the temperature dependence of the product $R_s M$ extracted at $\mu_0 H > 0$. Just below $T_{c,\mathrm{on}}$, the anomalous Hall signal $R_s M$ is small and slightly positive. As the temperature decreases, it exhibits an inflection near $T^*_{\mathrm{OHE}} = 155$ K, then drops rapidly and changes sign at $T^*_{\mathrm{AHE}} = 150$ K, becoming increasingly negative at lower temperatures. The small 5 K offset between the zero crossing of the ordinary Hall coefficient at $T^*_{\mathrm{OHE}} = 155$ K and the sign change of $R_s M$ at $T^*_{\mathrm{AHE}} = 150$ K suggests that both effects originate from the same underlying electronic reconstruction.

This sign reversal in anomalous Hall resistance reflects a change in the anomalous Hall coefficient $R_s$, which is proportional to the Berry curvature integrated over the occupied electronic states. The onset of ferromagnetic order introduces exchange splitting and thereby modifies the band structure near the Fermi level. In SRO, the anomalous Hall response is known to be highly sensitive to the spin-orbit-coupled multiband electronic structure and the resulting distribution of Berry curvature. Accordingly, the observed sign reversal of $R_s$ is consistent with a temperature-dependent redistribution of positive and negative Berry-curvature contributions near the Fermi level [2,3]. Notably, this occurs in the same temperature window where the ordinary Hall coefficient changes sign, indicating a concurrent shift in the dominant carrier character. This picture is consistent with recent studies on ultrathin SRO films, where Berry-curvature contributions are associated with multiple $t_{2g}$ -derived bands, particularly the $d_{xz}/d_{yz}$, while the change in dominant carrier character is linked to the $d_{xy}$ band near the Fermi level [12,13].

Figure 4 consolidates the full set of transport data into a temperature-resolved framework. All critical changes are confined to a narrow temperature window centered on the magnetic transition. At $T_{c,on} \simeq 165$ K, the resistivity anisotropy collapses and an AHE loop first appears, indicating the onset of long-range ferromagnetic order. A 5 K decrease brings the system to $T^*_{\mathrm{OHE}} = 155$ K, where the OHE coefficient $R_o$ crosses zero and the quadratic $MR$ term $C(T)$ peaks, indicating carrier compensation. Another 5 K step to $T^*_{\mathrm{AHE}} = 150$ K marks the sign reversal of $R_s M$, consistent with a temperature-dependent redistribution of Berry-curvature contributions near the Fermi level. At $T^0 = 145$ K, the linear MR along $[1\bar{1}0]$ changes sign and becomes positive, consistent with the emergence of a Berry-curvature-related linear magnetoresistance contribution [5]. Below 50 K, the system enters a conventional Fermi-liquid regime. Thus, several independently determined transport changes occur within a narrow temperature window around the ferromagnetic transition. Their close proximity is an important experimental observation, although it does not by itself establish a unique microscopic sequence.

**Discussion**

In the following, we focus on two key experimental observations from our transport measurements. (1) The temperature-dependent resistance $R(T)$, measured along the two orthogonal in-plane directions $[001]$ and $[1\bar{1}0]$, shows a pronounced anisotropy above the ferromagnetic ordering temperature $T_{c,\mathrm{on}}$, which becomes strongly suppressed below $T_{c,\mathrm{on}}$. This prompts the question: why does the zero-field $R(T)$ lose its directional dependence once long-range FM order sets in, despite the underlying crystallographic anisotropy remaining unchanged? (2) When a magnetic field is applied perpendicular to the film plane (along $[110]$), a clear directional dependence re-emerges in the magnetoresistance: the $MR$ behavior along $[001]$ and $[1\bar{1}0]$ differs significantly, even in the temperature range where the zero-field $R(T)$ curves nearly coincide. This raises the second question: what mechanism restores the directional sensitivity of charge transport in the presence of a magnetic field below $T_{c,\mathrm{on}}$?

In the semiclassical model, zero-field resistivity can be expressed as

$$\rho_i = \left[-e^2 \int \frac{d\boldsymbol{k}}{(2\pi)^3} \tau(\boldsymbol{k}) \frac{\partial f_0}{\partial \varepsilon} v_i^2(\boldsymbol{k})\right]^{-1}, \tag{3}$$

where $e$ is the electronic charge, the integration runs over all occupied $\boldsymbol{k}$-states, $\tau(\boldsymbol{k})$ is the momentum-dependent relaxation time, $v_i(\boldsymbol{k})$ is the component of the group velocity along $i$, and $f_0 = \frac{1}{1+e^{(\varepsilon-\varepsilon_F)/k_BT}}$ is the Fermi-Dirac distribution function [28]. Although the calculated Fermi-surface cross-sections perpendicular to $[001]$ and $[1\bar{1}0]$ remain non-identical (Figure S7 in sec 5. SI) throughout the temperature range, indicating different $v_{[001]}^2(\boldsymbol{k})$ and $v_{[1\bar{1}0]}^2(\boldsymbol{k})$ components due to orthorhombic symmetry, the observed directional dependence of $R(T)$ appears only above the ferromagnetic transition. This suggests that the impact of band anisotropy is strongly magnified in the paramagnetic regime because resistivity depends on the joint product $v_i^2(\boldsymbol{k})\tau(\boldsymbol{k})$ averaged over the Fermi surface. While $v_{[001]}^2(\boldsymbol{k})$ and $v_{[1\bar{1}0]}^2(\boldsymbol{k})$ remain different, the effective relaxation times $\tau(\boldsymbol{k})$ along the two directions become more similar in the ferromagnetic phase, thereby reducing the contrast between the corresponding transport integrals.

Within Moriya's spin-fluctuation framework, $\tau(\boldsymbol{k})$ can acquire momentum-dependence because electron-scattering is governed by the momentum- and frequency-dependent magnetic fluctuation spectrum [10]. In systems with anisotropic spin fluctuations, different regions of the Fermi surface experience different scattering strengths, which produces direction-dependent resistivity even in nearly tetragonal materials. This mechanism is well established experimentally. For example, detwinned $Ba(Fe_{1-x}Co_x)_2As_2$ develops a pronounced in-plane resistivity anisotropy that originates from electronically driven nematic fluctuations rather than structural distortions [29]. Subsequent theoretical work demonstrated how anisotropic spin fluctuations generate a direction-dependent relaxation time and quantitatively reproduce such transport anisotropy [30]. In our case, below $T_{c,on}$, the anisotropy in $R(T)$ collapses as spin-disorder scattering is suppressed. At lower temperatures, the resistivity is governed predominantly by electron-magnon and electron-electron scattering, which account for the observed $T^{3/2}$ and $T^2$ dependences, respectively [20], [22]. We do not assume that these scattering mechanisms are intrinsically isotropic. Rather, the experimentally observed convergence of the two $R(T)$ curves indicate that the corresponding Fermi-surface-averaged transport integrals become nearly equal in the ferromagnetic state. In other words, the directional contrast associated with the effective relaxation time is strongly reduced below $T_{c,\mathrm{on}}$, even though the underlying band-velocity anisotropy remains finite. This supports our interpretation that the pronounced high-temperature anisotropy originates primarily from the momentum-dependent spin-fluctuation scattering in the paramagnetic phase.

An alternative explanation for the vanishing anisotropy in the FM phase could be a random distribution of magnetic domains at zero field, which would lead to isotropic scattering as charge carriers encounter domain walls in all directions. However, we negate this possibility by showing that a (110)-SRO film grown on (110)-$NdGaO_3$ substrate exhibits no influence on $R(T)$ anisotropy whether measured in zero and 1 T field. Because a 1 T field is sufficient to align all magnetic domains (see Figure S6 in SI sec. 4), the absence of any field dependence confirms the domain averaging does not account for the disappearance of anisotropy.

While the zero-field resistivity becomes nearly isotropic below $T_{c,\mathrm{on}}$, the linear component of the magnetoresistance reveals a sharp directional contrast. Along $[001]$, it remains negative and saturates; along $[1\bar{1}0]$, it reverses sign near $T^0 = 145$ K and becomes positive at lower temperatures, as shown in Figure 2(c). This divergence occurs well below the carrier compensation point $T^*_{\mathrm{OHE}} = 155$ K, where the ordinary Hall coefficient $R_o$ crosses zero, and just below $T^*_{\mathrm{AHE}} = 150$ K, where the anomalous Hall signal $R_s M$ changes sign, as summarized in Figure 4.

The sign changes in the ordinary and anomalous Hall components occur in close proximity across the ferromagnetic transition. The reversal of the ordinary Hall coefficient $R_0$ at $T^*_{\mathrm{OHE}} = 155$ K marks a crossover in the dominant Hall response from hole-like to electron-like, whereas the subsequent sign change of $R_s M$ at $T^*_{\mathrm{AHE}} = 150$ K reflects a change in the anomalous Hall coefficient $R_s$. In the semiclassical framework of Zhang *et al.* [28], the anomalous Hall resistivity is expressed as

$$\rho^A_{yx} = \left[\frac{e^2}{\hbar}\int \frac{d\boldsymbol{k}}{(2\pi)^3}\boldsymbol{\Omega}_z(\boldsymbol{k}) f_0\right]^{-1}, \quad (4)$$

where the subscript $z$ denotes the component of the Berry curvature $(\Omega)$ parallel to the applied magnetic field, which in our geometry corresponds to the $[110]$ direction. Our first-principles calculations based on the experimental lattice parameters of the SRO film reveal sharp, monopole-like peaks in $\Omega_{[110]}(\boldsymbol{k})$ of both positive and negative polarity near avoided band crossings in the ferromagnetic phase (Figure S8 in sec 6. SI). These positive and negative Berry-curvature peaks contribute strongly to the integral in eq. (4), making the anomalous Hall response highly sensitive to the changes in the electronic structure near the Fermi level. The sequential switching of $R_o$ and $R_s M$ is consistent with temperature-dependent redistribution of Berry curvature accompanying the transition from hole- to electron-dominated transport, consistent with our DFT results showing that small Fermi-level shifts across the magnetic transition reverse the relative weights of positive and negative Berry-curvature monopoles near the Fermi surface.

The directional dependence of the linear magnetoresistance originates from the Berry-phase modification of carrier dynamics, described by the semiclassical equation of motion [28]:

$$\dot{\boldsymbol{r}} = \frac{1}{\hbar}\frac{\partial \varepsilon_n(\boldsymbol{k})}{\partial \boldsymbol{k}} + \frac{e}{\hbar}\boldsymbol{E} \times \boldsymbol{\Omega}_n(\boldsymbol{k}) + \frac{e}{\hbar}(\boldsymbol{v}_n \times \boldsymbol{B}) \times \boldsymbol{\Omega}_n(\boldsymbol{k}). \quad (5)$$

The second term in eq. (5) produces the anomalous transverse velocity, while the third term couples the Berry curvature to the magnetic field and generates the linear component of magnetoresistance. The last term acts opposite to the band velocity, effectively slowing carrier motion when a magnetic field is applied and giving rise to a linear increase in resistivity [28]. In our measurement geometry, the magnetic field is applied along $[110]$, and the relevant Berry-curvature component is $\Omega_{[110]}(\boldsymbol{k})$. As detailed in SI, the linear Berry-curvature-related contribution to the longitudinal transport is weighted by $v_i^2(k)\Omega_{[110]}(k)$. Therefore, even for the same field direction, its magnitude can differ strongly between the two crystallographic current directions because of their different band velocities. The DFT calculations show that the band velocities are larger along $[1\bar{1}0]$ near the $k$-points where $|\boldsymbol{\Omega}(\boldsymbol{k})|$ exhibits sharp monopole-like peaks, which enhances the Berry-curvature contribution to transport in this direction (Figure S9 in sec 6. SI). This provides a possible microscopic origin for the positive linear MR observed along $[1\bar{1}0]$. Along $[001]$, the smaller carrier velocity near the same Berry-curvature hotspots results in a weaker Berry-curvature-related linear contribution, which is insufficient to overcome the negative MR contribution within our measured temperature range. We emphasize that the present calculations provide a qualitative microscopic basis for the observed directional transport, while a complete quantitative description would require a more comprehensive transport calculation. The directional MR sign reversal is reproduced in an additional 25 nm SRO/DSO film (Figure S12 in SI). Furthermore, linear four-probe measurements on the primary 50 nm SRO/DSO film reproduce the MR anisotropy observed with the van der Pauw configuration, confirming that the effect is not an artifact of the measurement geometry (Figure S13, Supporting Information).

Within the same framework based on Berry curvature and band velocity, no directional dependence is expected in the Hall response, for both ordinary and anomalous components (see sec. 7 in SI). Consistent with this, the measured Hall resistance shows no discernible anisotropy between the $[001]$ and $[1\bar{1}0]$ directions (Figures S10 and S11 in sec. 8 SI).

**Conclusion**

We have investigated transport anisotropy in an epitaxial (110)-oriented $SrRuO_3$ thin film grown on $DyScO_3$, revealing a tightly coupled sequence of magnetic, electronic, and topological crossovers across the ferromagnetic transition. In the paramagnetic bad-metal regime, resistivity along $[001]$ and $[1\bar{1}0]$ differs markedly, consistent with anisotropic spin-fluctuation scattering. Below the ferromagnetic ordering temperature $T_{c,\text{on}} = 165\,\text{K}$, this anisotropy is significantly reduced, as directionally averaged electron-magnon and electron-electron scattering dominate.

However, the expected isotropic magnetoresistance response does not emerge; under an applied magnetic field, the linear $MR$ becomes sharply direction dependent. While it remains negative along $[001]$, it changes sign and becomes positive along $[1\bar{1}0]$ below $T^0 = 145$ K, growing in magnitude at lower temperatures. These features unfold within a narrow 20 K window from $T_{c,\text{on}}$, together with carrier compensation at $T^*_{\text{OHE}} = 155$ K, a sign reversal of anomalous Hall contribution at $T^*_{\text{AHE}} = 150$ K consistent with Berry-curvature redistribution, and the directional MR reversal at $T^0$. The close proximity of these characteristic temperatures is consistent with an evolution of the spin-orbit-coupled multiband electronic structure across the ferromagnetic transition, accompanied by a redistribution of Berry-curvature contributions near the Fermi level [2], [3].

Our results show that $SrRuO_3$ hosts two distinct sources of magnetotransport anisotropy: a field-induced contribution below $T_{c,on}$, associated with anisotropic band velocities near Berry-curvature-rich regions of the electronic structure, a zero-field anisotropy above $T_{c,on}$ governed predominantly by spin-fluctuation scattering. This duality makes the region near the ferromagnetic transition a powerful setting for exploring symmetry-sensitive and Berry-curvature-related magnetotransport.

**Experimental Section/Methods**

Details of the sample growth, structural characterization (Figures S1-S3 in SI), magnetotransport measurements (Figure S4 in SI), and first-principles calculations are provided in section 1 of the SI.

**Acknowledgements**

K.S. acknowledges the financial support from the BRNS under YSRP, the DST INSPIRE Faculty Fellowship, Core Research Grant from the SERB, India. S.B. acknowledges PhD fellowship from the CSIR India. We also acknowledge the SMILE-13 project. All authors acknowledge the magneto-transport facility at the Institute Instrumentation Centre of IIT Roorkee.

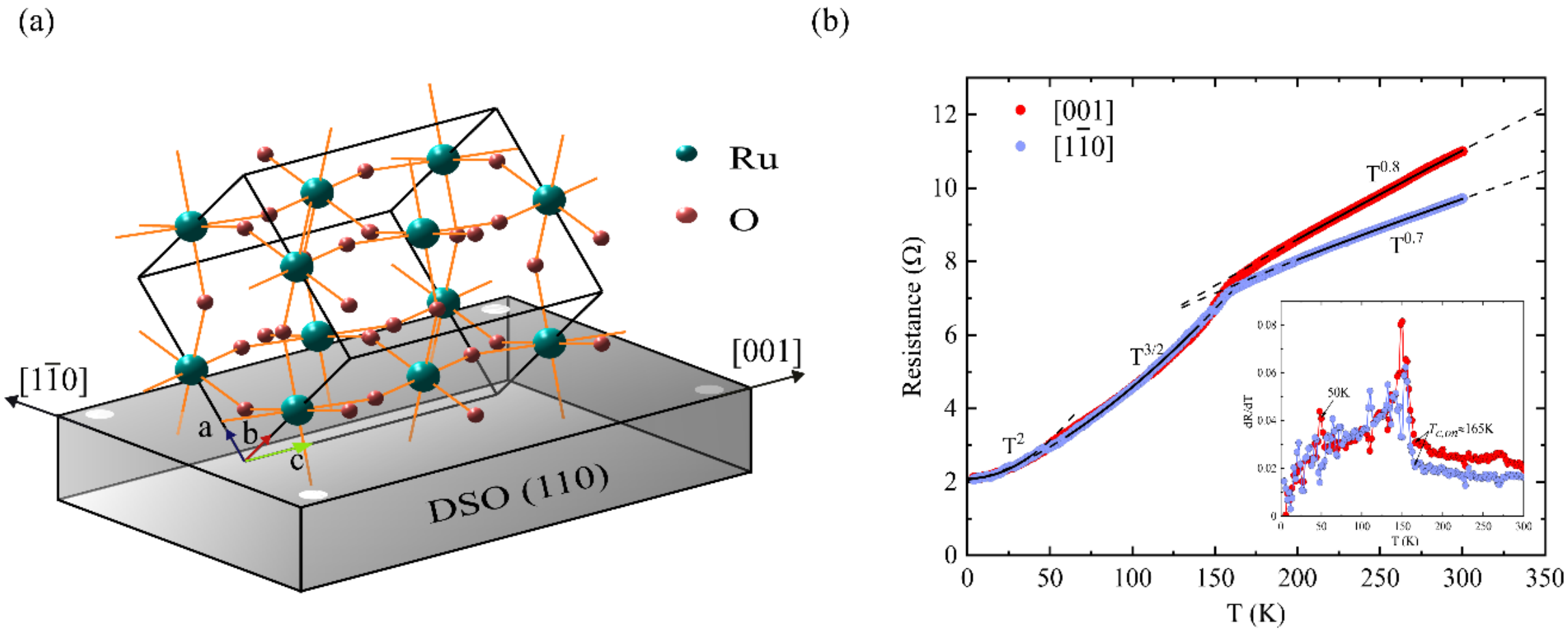

**Figure 1.** (a) Schematic diagram of the epitaxial (110)-oriented $SrRuO_3$ film, showing the (110)-oriented crystallographic unit cell of $SrRuO_3$ aligned on top of the (110)-oriented $DyScO_3$ substrate. The long $c$-axis lies in-plane along [001]. The $a$- and $b$-axes are rotated $45^o$ relative to the film normal. (b) Zero-field resistance ($R$) versus temperature ($T$) along two orthogonal in-plane directions [001] (red symbols) and [1$\bar{1}$0] (blue symbols). The solid lines through the data represent the best fits to $R(T)$. As mentioned in the figure, different $R-T$ regions are fitted with different exponents of $T$. The inset in (b) shows the derivative of resistance with respect to temperature, where $T_{c,on}$ represents the ferromagnetic transition temperature.

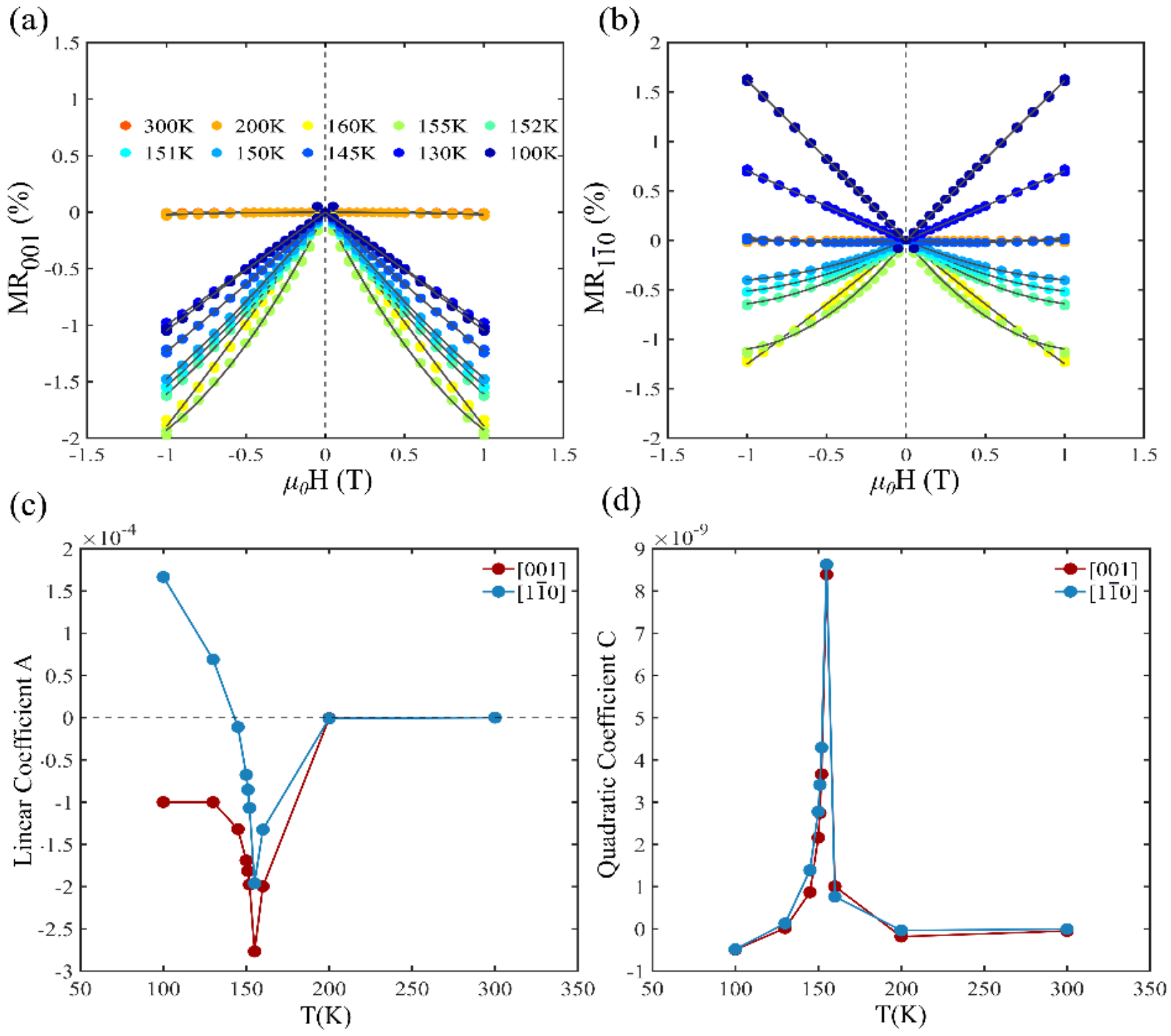

**Figure 2.** Magnetic field $(\mu_0 H)$ dependence of magnetoresistance (MR, symbols) along two in-plane directions (a) $[001]$, and (b) $[1\bar{1}0]$. The solid lines through the data are the best fit according to model $MR(\mu_0 H) = A(T)|\mu_0 H| + C(T)(\mu_0 H)^2$. (c) The obtained linear coefficient $A(T)$ as a function of temperature for two in-plane directions. (d) The same for the quadratic coefficient $C(T)$.

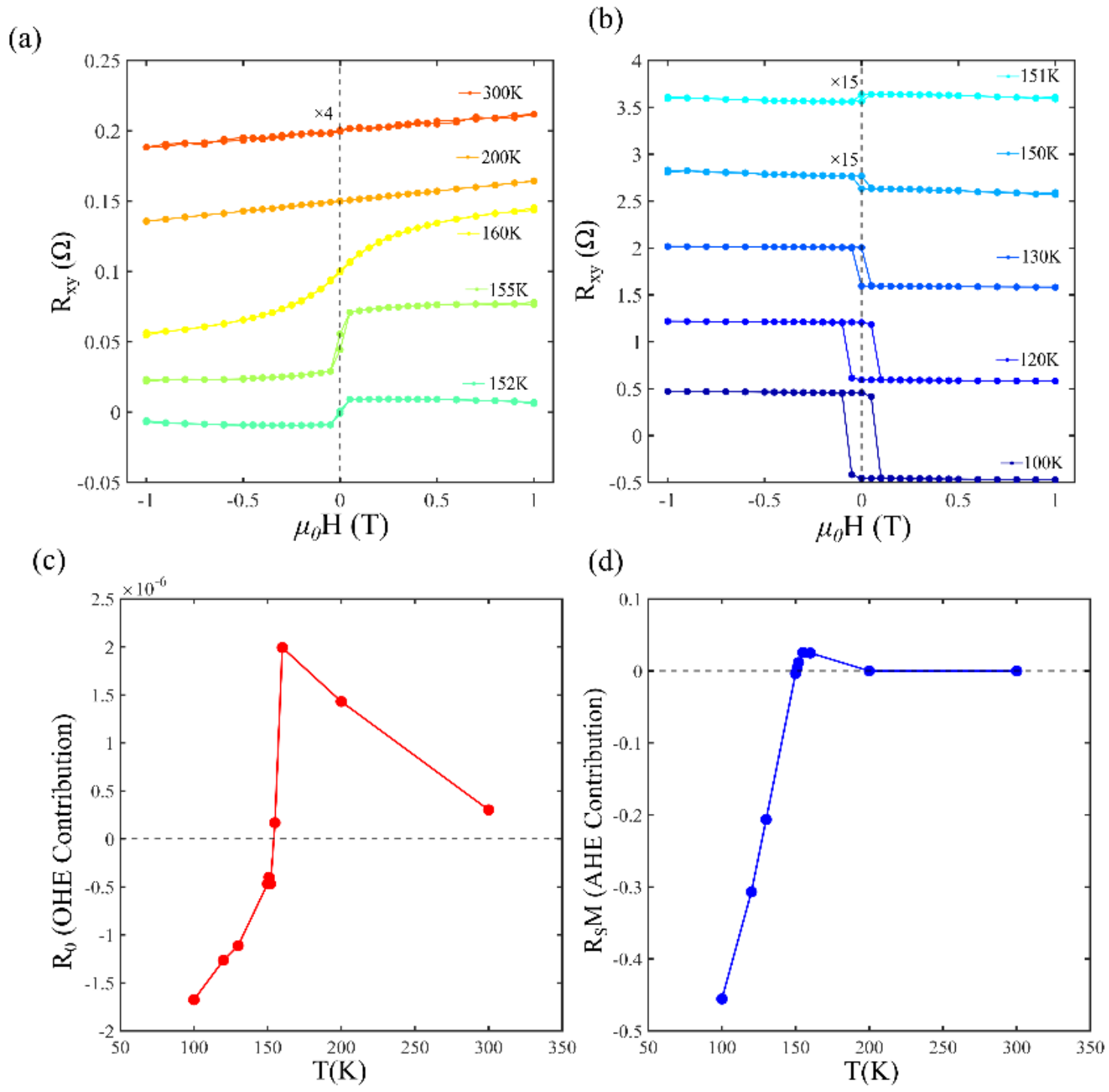

**Figure 3.** (a) and (b) Hall resistance as a function of applied magnetic field at several temperatures. (c) Ordinary Hall coefficient, whose sign indicates change in the nature of dominant charge carriers, i.e. holes or electrons. (d) Total anomalous Hall effect part (Rs*M) as a function of temperature.

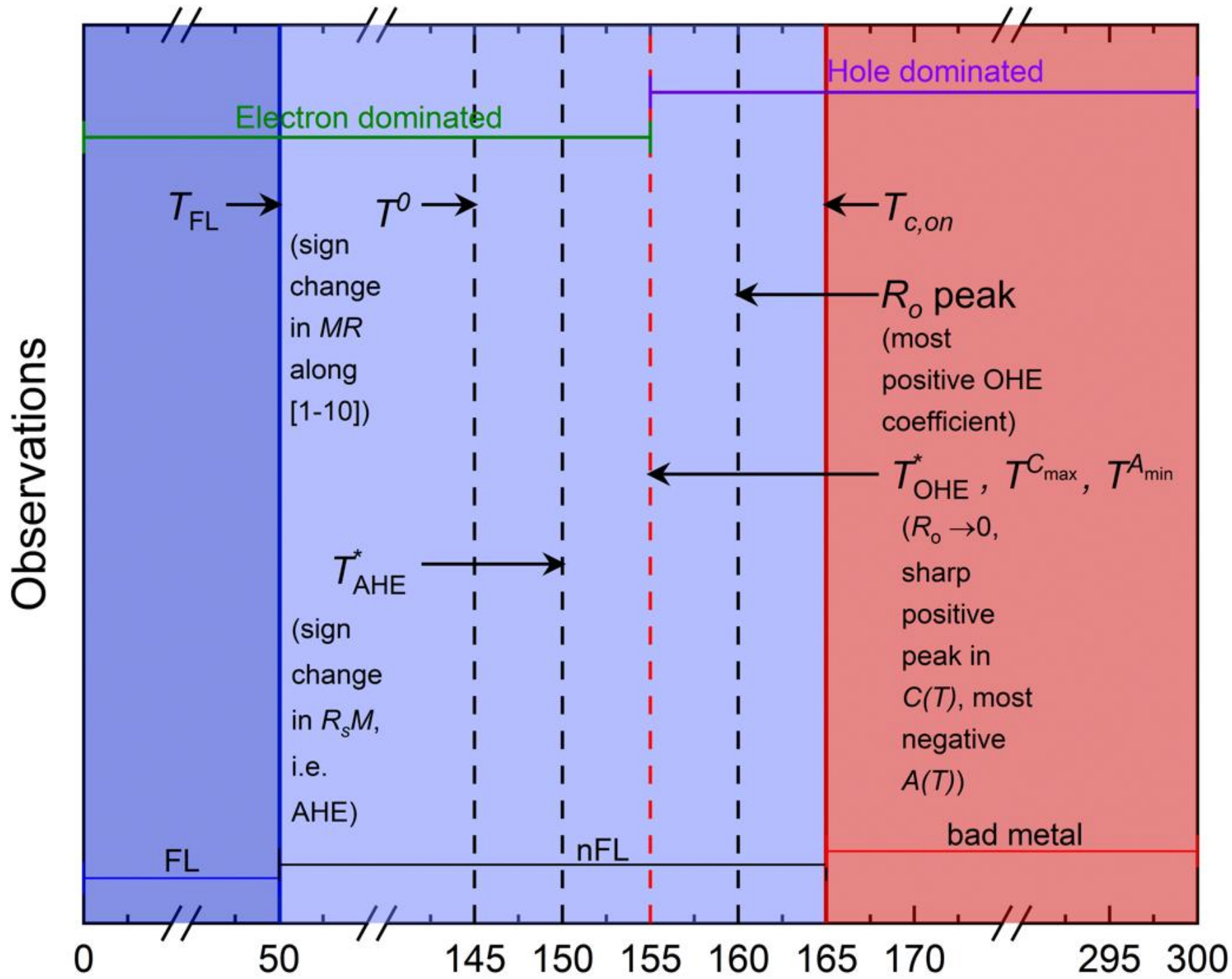

**Figure 4.** Summary of key transport crossovers in (110)-oriented 50 nm thick SRO. Vertical lines mark characteristic temperatures extracted from MR and Hall data. Coloured regions indicate Fermi-liquid (FL), non-Fermi-liquid (nFL), and paramagnetic bad-metal regimes. Carrier type is inferred from the sign of the ordinary Hall Effect coefficient.

Supplementary Information

for

**Suppression and reemergence of directional transport across the ferromagnetic transition in $SrRuO_3$**

S. Bhatia *et al.*

## 1. Materials and methods

### 1.1. Thin film deposition

(110) oriented perovskite SRO films were grown on $DyScO_3$ (110) substrates using a pulsed laser deposition system with a KrF excimer laser ( $\lambda$ = 248 nm). The film was deposited at a substrate temperature $T_S$ = 700˚C. Stochiometric deposition of SRO was obtained for an oxygen partial pressure of 0.1 mbar. After deposition, in-situ annealing was performed for 30 min at 500˚C and an increased pressure of 500 mbar. Subsequently, the sample was cooled down to room temperature.

### 1.2. Structural characterization

The symmetry and lattice parameter of the SRO film grown on DSO (110) were analyzed by XRD using a standard Bruker D8 Davinci diffractometer equipped with Cu-$K\alpha$ radiation. In the following, the orthorhombic symmetry is described by the space group *Pbnm*, which corresponds to the point group $D_{2h}$. Figure S1 shows higher-order symmetric reflections from the lattice planes perpendicular to the surface normal.

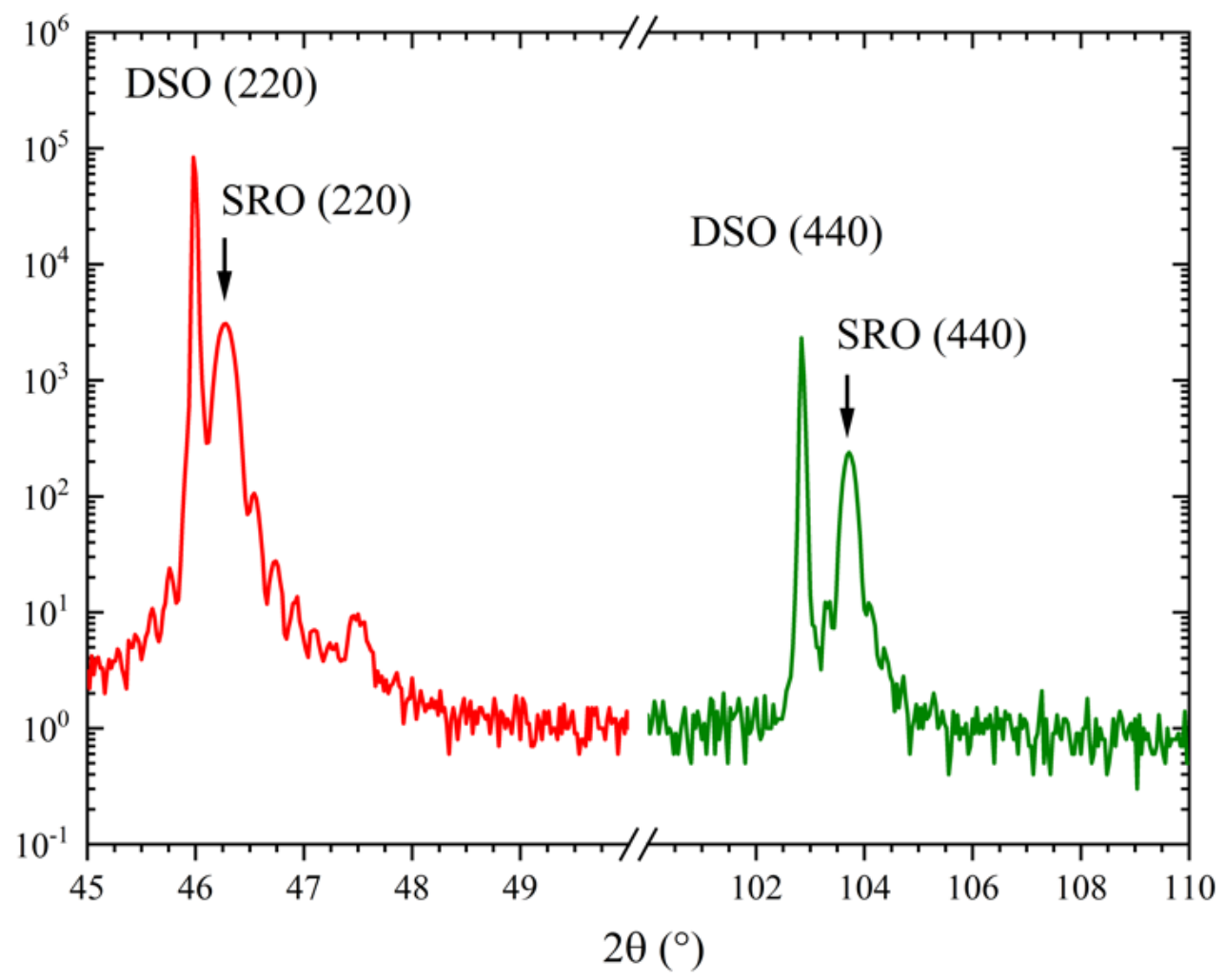

**Figure S1.** Symmetric x-ray diffraction of SRO films deposited on (110) oriented DSO. Film peak position corresponding to (220) and (440) are indicated by arrow.

The pseudo-cubic axes $a$, $b$, and $c$ are chosen to be parallel to the orthorhombic axes of the DSO substrate $[\mathbf{1\bar{1}0}]$, $[\mathbf{001}]$, and $[\mathbf{110}]$. To determine orthorhombic distortion and in-plane lattice parameters, we carried out reciprocal space mapping (RSM) on various asymmetric reflections corresponding to the orthorhombic SRO $(\mathbf{444})$, $(\mathbf{260})$, $(\mathbf{44\bar{4}})$, and $(\mathbf{620})$ planes. Figure S2 shows the reciprocal space map around these asymmetric SRO peaks. The SRO film displays clear orthorhombic distortion with experimental (bulk) lattice parameters $a$ = 5.532 Å (5.530 Å), $b$ = 5.568 Å (5.567 Å), and $c$ = 7.957 Å (7.845 Å ) [1].

Besides the distinct orthorhombic distortion ($a \neq b$), the $c$-axis is slightly elongated with respect to the bulk. The small tensile strain along the $c$-axis (~1.4%) and enhanced orthorhombic distortion are caused by the epitaxial growth on DSO ($a$ = 5.443 Å, $b$ = 5.719 Å, and $c$ = 7.907 Å).

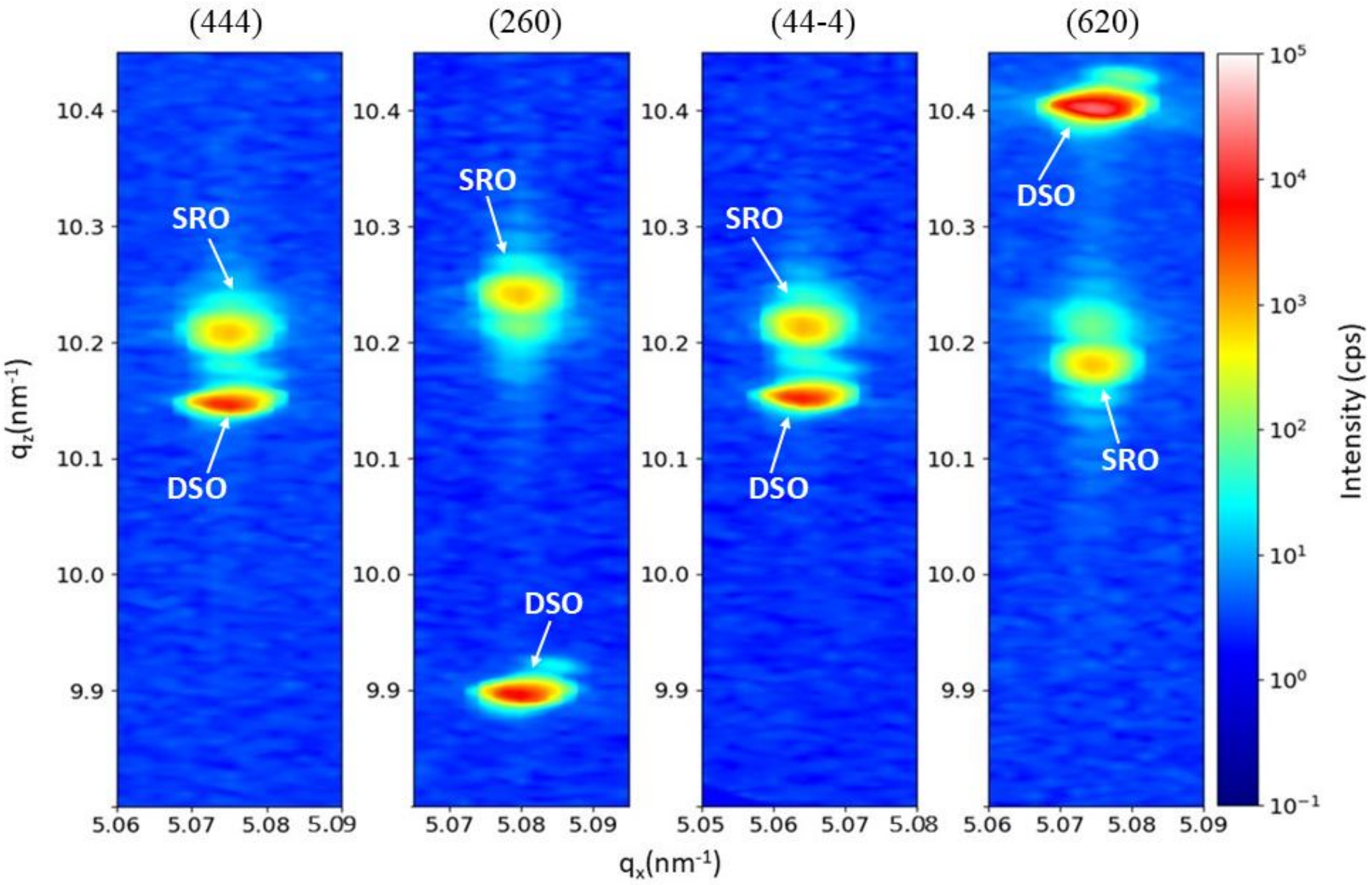

**Figure S2.** Reciprocal space map around asymmetric reflections of SRO film on (110) DSO.

From the half-integer pseudo-cubic (pc) reflections the $RuO_6$ octahedral tilt/rotation pattern can be determined. According to the Glazer notation, the pseudo-cubic reflections $\langle\mathbf{131}\rangle$ indicate $c^-$ and $a^-$ rotation, the $\langle\mathbf{311}\rangle$ indicate $c^-$ and $b^-$, and the $\langle\mathbf{113}\rangle$ indicate $b^-$ and $a^-$. For clockwise octahedral rotation the $\langle\mathbf{213}\rangle$ show $a^+$, the $\langle\mathbf{123}\rangle$ show $b^+$ and the $\langle\mathbf{312}\rangle$ indicates $c^+$. As shown in Figure S3, the orthorhombic $(\mathbf{112})$ reflection corresponds to the half-integer pseudo-cubic (pc) reflection $(\mathbf{213})$ and indicates clockwise rotation about the pc *a*-axis ($a^+$), *i.e.*, the orthorhombic *c*-axis. The orthorhombic $(\mathbf{121})$ and $(\mathbf{211})$ reflections correspond to the pc $(\mathbf{1\bar{1}3})$ and $(\mathbf{1\bar{3}1})$ and indicate $b^-$ or $a^-$ and $c^-$or $a^-$ rotation, respectively. Since rotation about pc *a*-axis is clockwise (+) these reflections are caused by $b^-$ and $c^-$ rotation, respectively. To prove the absence of clockwise rotation $b^+$ and $c^+$ we have looked for half-integer reflections of the orthorhombic $(\mathbf{152})$ and $(\mathbf{136})$ reflections, respectively. Intensity at the expected 2θ position (see arrows in the figure) is clearly absent. Hence, a rotation pattern of pc ($a^+b^-c^-$) or with respect to the orthorhombic unit cell ($a^-b^-c^+$) is deduced. This is the same rotation pattern as observed for bulk SRO.

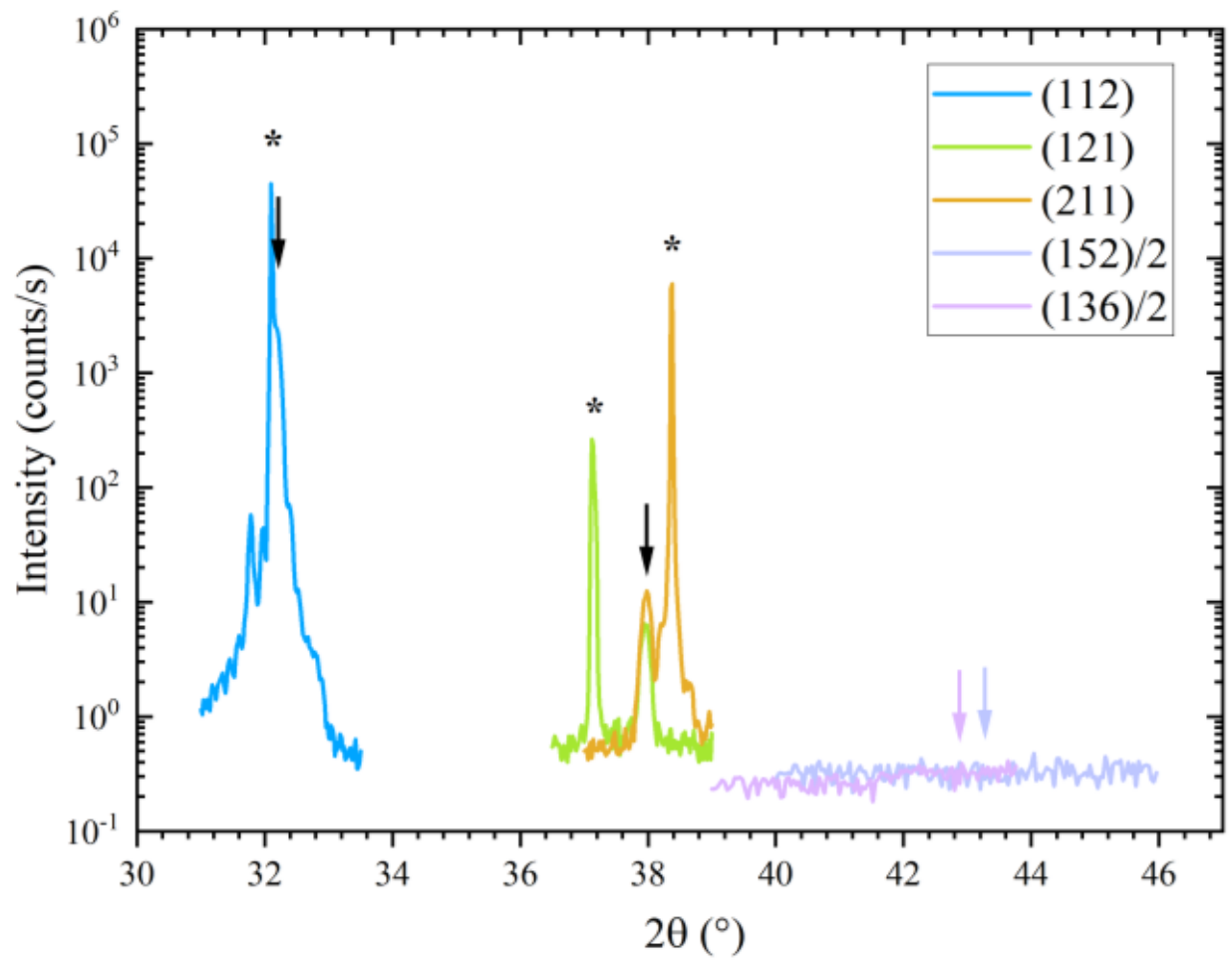

**Figure S3.** Orthorhombic reflections corresponding to half-integer pseudo cubic reflections of SRO on DSO.

**1.3. Resistance measurement**

Figure S4 schematically illustrates the Van der Pauw geometry used for in-plane resistance measurements. We employed a Keithley switch (model 3706 A) to systematically apply DC bias currents $\boldsymbol{I}$ using Keithley current source (model 6221) along four in-plane edges of the sample and measured the corresponding voltages $\boldsymbol{V}$ using Keithley voltmeter (model 2182A). From these data, we determined resistances along specific crystallographic directions, such as $\boldsymbol{R}_{[\mathbf{001}]} = \boldsymbol{V}_{[\mathbf{001}]}/\boldsymbol{I}_{[\mathbf{001}]}$ and $\boldsymbol{R}_{[\mathbf{1\bar{1}0}]} = \boldsymbol{V}_{[\mathbf{1\bar{1}0}]}/\boldsymbol{I}_{[\mathbf{1\bar{1}0}]}$. Each reported resistance is the average of four measurements, which involved reversing the current polarity and selecting the relevant edges parallel to $[\mathbf{001}]$ and $[\mathbf{1\bar{1}0}]$.

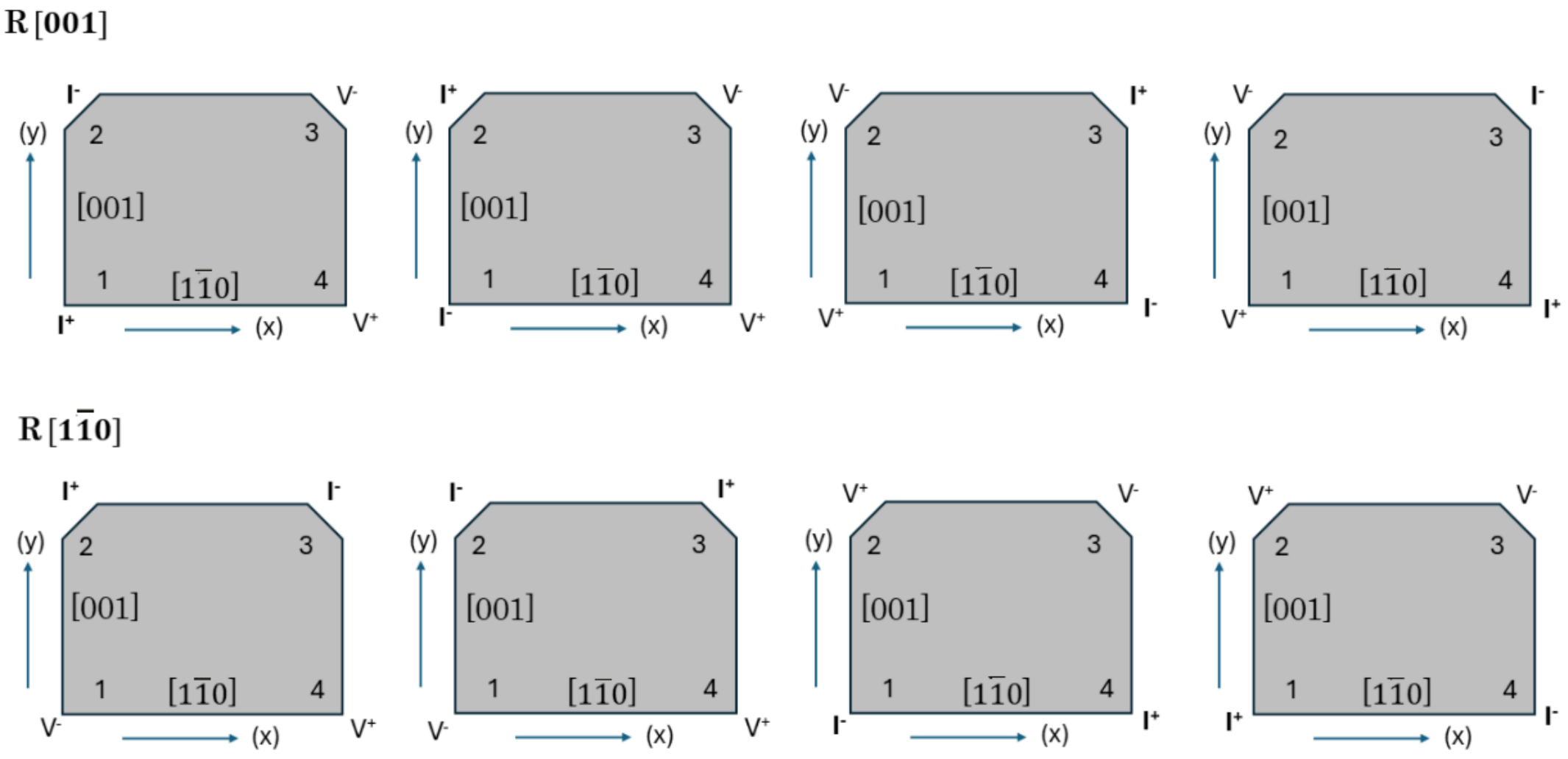

**Figure S4.** Van der Pauw geometry used for in-plane resistance measurements along [001] and [1$\bar{1}$0].

## 1.4. Symmetrization of magnetoresistance data

Symmetrizing magnetoresistance (MR) data is done to eliminate contributions from the Hall voltage and experimental asymmetries. It involves averaging the resistance values measured under positive and negative magnetic fields [2].

$$\boldsymbol{R_{xx}\,(H) = \frac{R_{xx}(H) + R_{xx}(-H)}{2}} \qquad (1)$$

All MR values reported in the main text are based on this symmetrized data.

## 1.5. Anti-Symmetrization of Hall Resistance data:

To isolate the true Hall resistivity and eliminate spurious contributions from contact misalignment or longitudinal resistivity mixing, the transverse voltage data were antisymmetrized using [2]:

$$\boldsymbol{R_{xy}\,(H) = \frac{R_{xy}(H) - R_{xy}(-H)}{2}} \qquad (2)$$

All Hall resistivity values and Hall coefficient analyses in this work are based on these antisymmetrized data.

## 1.6. First-principles DFT calculations

We performed the first-principles DFT calculations using Vienna ab-initio Simulation Package(VASP) [3]. We have used the Perdew-Burke-Ernzerhof generalized-gradient approximation (PBE-GGA) as the exchange-correlation functional [4]. We included the Coulomb correlation for Ru d-states with $U_{eff}$ = U - J = 0.6 eV (where U is Coulomb correlation and J is Hund's exchange) and the spin-orbit coupling within GGA+U+SO approximation. The $U_{eff}$ value used here was taken from a previous study where the authors found a good agreement with Photoemission spectroscopy (PES) data [5]. The crystal structure was optimized until the Hellmann-Feynman forces converged to 0.01 eV/Å or less, maintaining the lattice parameters at their respective experimental values. In order to match the experimental directions, we considered a $\sqrt{2} \times \sqrt{2} \times 1$ supercell. The energy convergence criterion for self-consistent calculations was set to $10^{-6}$ eV. A Γ centered 7×7×7 Monkhorst Pack grid was chosen for structural optimization. We used a 3×3×3 k-points grid to construct the Maximally Localized Wannier Functions (MLWFs) and to calculate the Berry curvature with spin-orbit coupling [6,7]. We have considered non-magnetic and ferromagnetic structures to explain the behaviour above and below the ferromagnetic transition observed in our experimental measurements. The ferromagnetic structure is found to be the ground state with magnetic moments of 2 $\mu_B$ per Ru ion consistent with the experimental values [8].

**2. Analysis of the MR data**

From the semiclassical two-band model, the electrical resistivity of materials with both electron-type and hole-type carriers under a vertical magnetic field is given by [9],[10]:

$$\boldsymbol{\rho(\mu_0 H) = \frac{(n_e\mu_e + n_h\mu_h) + (n_e\mu_h + n_h\mu_e)\mu_e\mu_h(\mu_0 H)^2}{e(n_e\mu_e + n_h\mu_h)^2 + e(\mu_e\mu_h)^2(n_e - n_h)^2(\mu_0 H)^2}} \qquad (3)$$

where $\boldsymbol{n_e}$ ($\boldsymbol{n_h}$) denotes the electron (hole) concentration, $\boldsymbol{\mu_e}$ ($\boldsymbol{\mu_h}$) represents the mobility of electrons (holes), e signifies the elementary charge of the electron, and $\boldsymbol{\mu_0 H}$ indicates the magnetic field. Subsequently, the Magnetoresistance, which denotes the variation in electrical resistance due to the influence of a magnetic field, produces

$$\boldsymbol{MR = \frac{R(\mu_0 H) - R(0)}{R(0)} \times 100\ \%} \qquad (4)$$

$$\boldsymbol{MR = \frac{(n_e\mu_e n_h\mu_h)(\mu_e + \mu_h)^2(\mu_0 H)^2}{(n_e\mu_e + n_h\mu_h)^2 + (n_e - n_h)^2(\mu_0 H)^2}} \qquad (5)$$

When charge compensation is achieved ($\boldsymbol{n_e} = \boldsymbol{n_h}$), the magnetoresistance exhibits a quadratic relationship with the magnetic field.

$$\boldsymbol{MR = \mu_e\mu_h(\mu_0 H)^2} \qquad (6)$$

Therefore, the quadratic coefficient in the main text expression $\boldsymbol{C(T)}$ is identified as:

$$C(T) = \mu_e \mu_h \tag{7}$$

## 3. Anisotropy in R(T) along two in-plane directions

Figure S5 compares the temperature-dependent in-plane resistance anisotropy of our SRO film grown on DSO with previously reported data from Klein et al. [11] and Ziese et al. [12] on SRO films. In the three datasets, strong direction dependence has been observed above the ferromagnetic transition. Klein *et al.* (panel b) most closely resemble our results and the temperature dependence of the anisotropy. In contrast, Ziese *et al.* (panel c) report a persistent anisotropy across the entire temperature range, including within the ferromagnetic state.

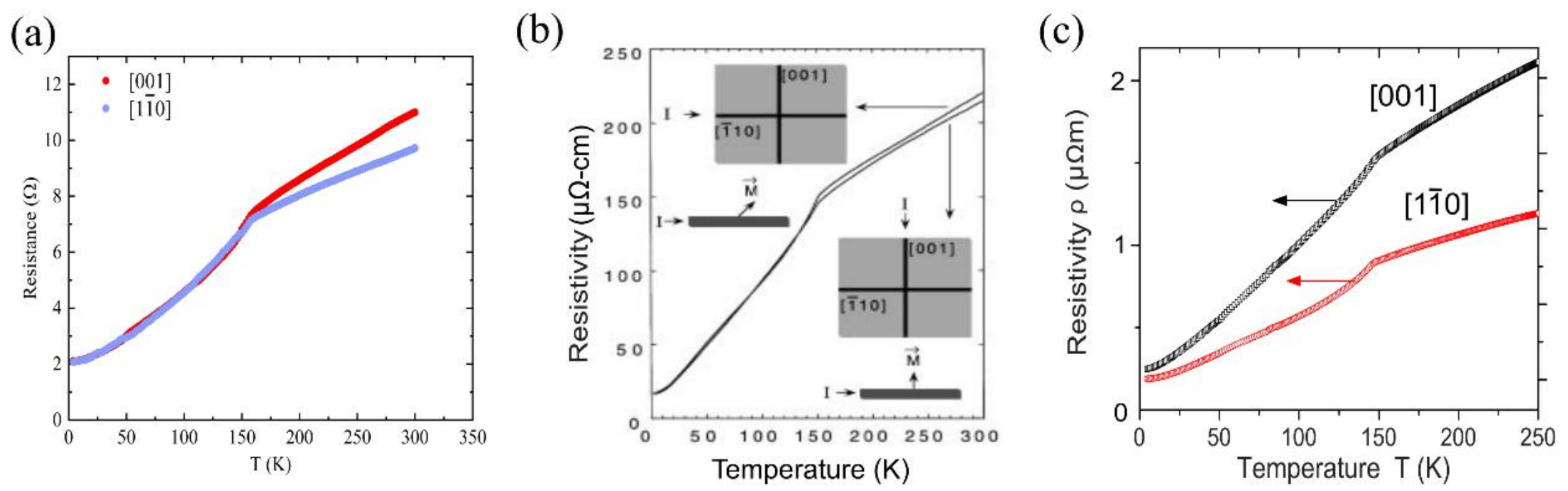

**Figure S5.** Temperature-dependent in-plane resistance of SRO films (a) Our work, (b) Klein et al, (c) Ziese et al.

## 4. R(T) on SRO grown on (110)-NGO substrate

Figure S6. Resistance *R* versus temperature *T* along two orthogonal in-plane directions [001] and 110 of SRO on NGO substrate at zero field and at 1T. In this film, we compared the temperature-dependent resistivity along orthogonal in-plane directions under zero field and under an applied magnetic field of 1 T. The R(T) anisotropy remains unchanged in both

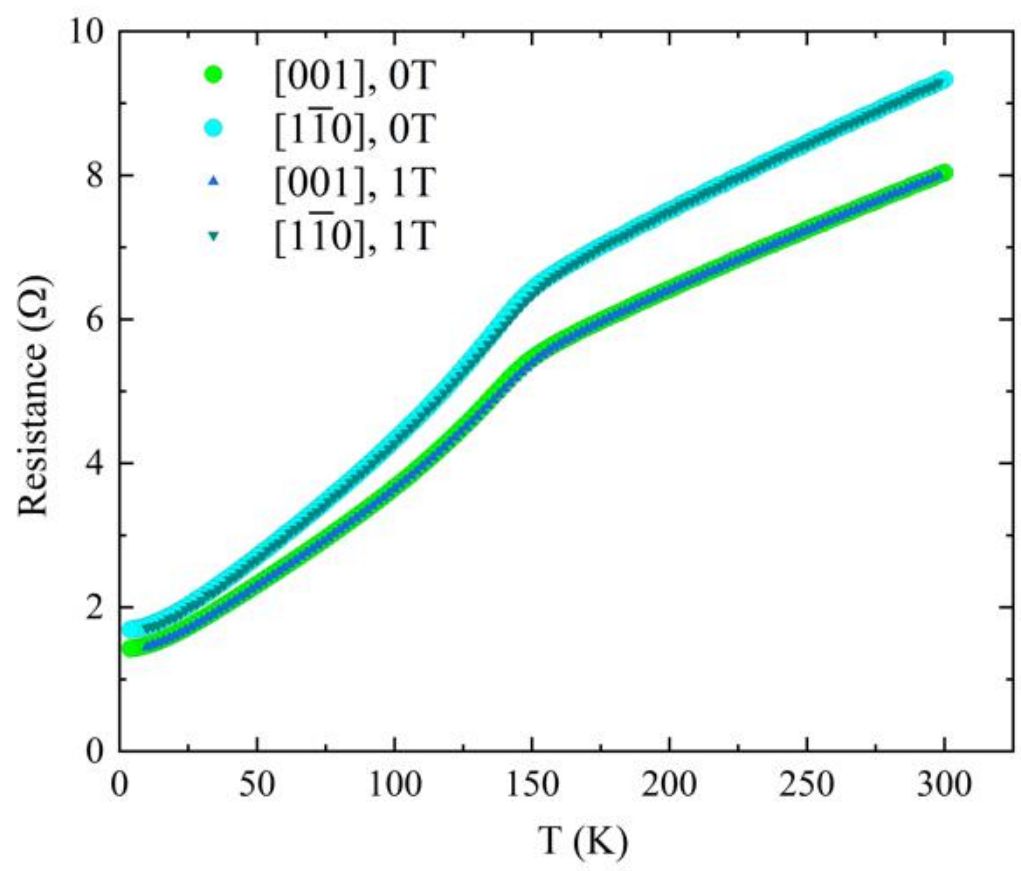

**Figure S6.** Resistance $(R)$ versus temperature $(T)$ along two orthogonal in-plane directions $[001]$ and $[1\bar{1}0]$ of SRO on NGO substrate at zero field and at 1T.

conditions, indicating that magnetic domain alignment does not affect the observed anisotropy. These observations rule out domain averaging as the cause of the vanishing anisotropy in our SRO film on DSO.

## 5. Fermi Surface

In Figure S7, we present the Fermi surface cross sections perpendicular to $[\mathbf{1\bar{1}0}]$ and $[\mathbf{001}]$ directions for non-magnetic and ferromagnetic phases. A clear anisotropy is observed between $[\mathbf{1\bar{1}0}]$ and $[\mathbf{001}]$ directions in both non-magnetic and ferromagnetic phases.

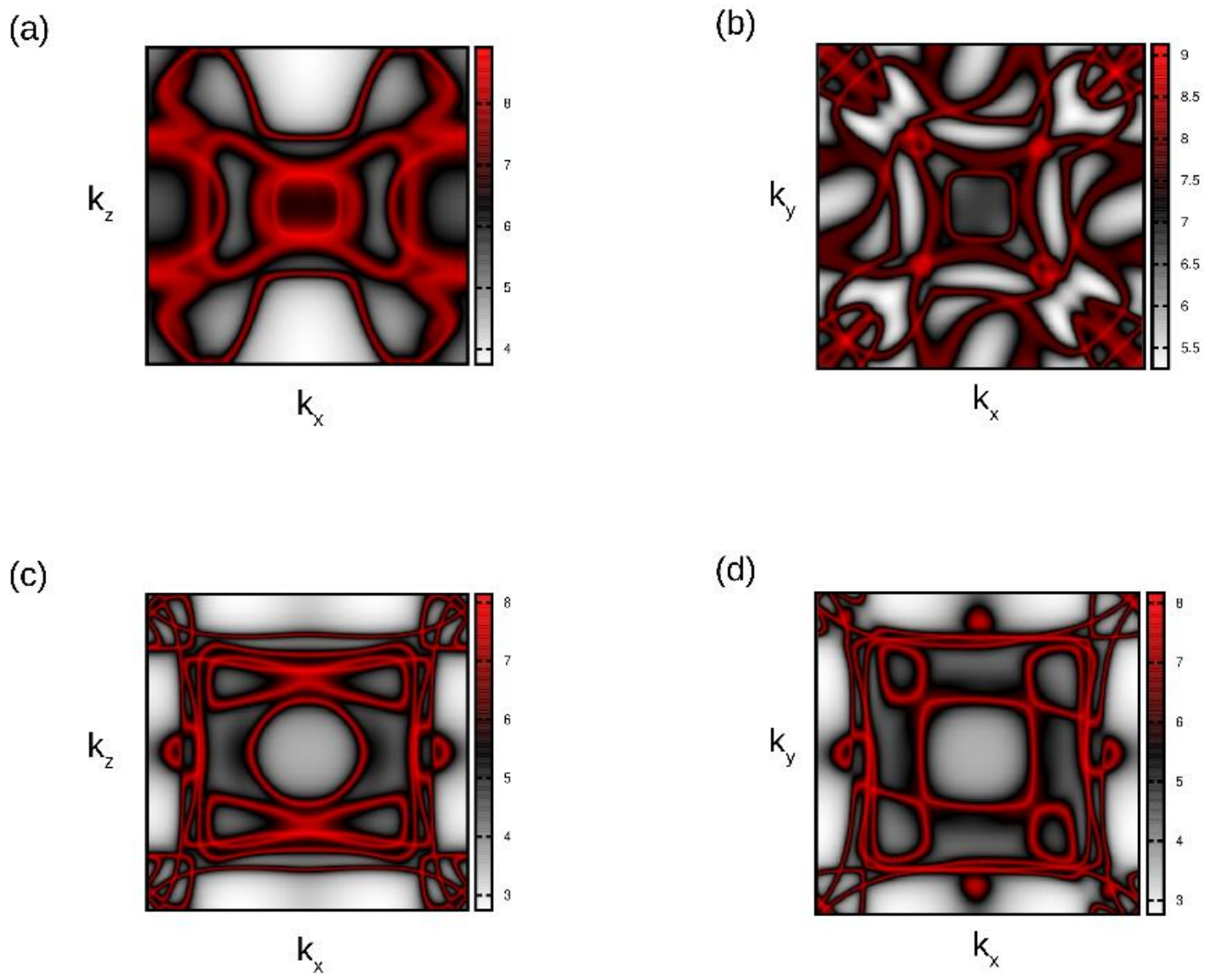

**Figure S7**. The Fermi surface cross sections perpendicular to (a) $[1\bar{1}0]$ and (b) $[001]$ directions for non-magnetic phase and (c) $[1\bar{1}0]$ and (d) $[001]$ directions for ferromagnetic phase.

## 6. Calculation of Berry curvature

The calculated Berry curvatures is shown in Figure S8. We can see huge Delta-like peaks with extremely large Berry curvature values (of the order of $10^6$ ($Å^2$)). Such huge values of Berry curvature (known as Berry curvature monopoles) have been seen in other topological materials where the time-reversal symmetry is broken. Therefore, our system, due to the strong topological character, is expected to show linear positive magnetoresistance (LPMR) in its ferromagnetic state [13].

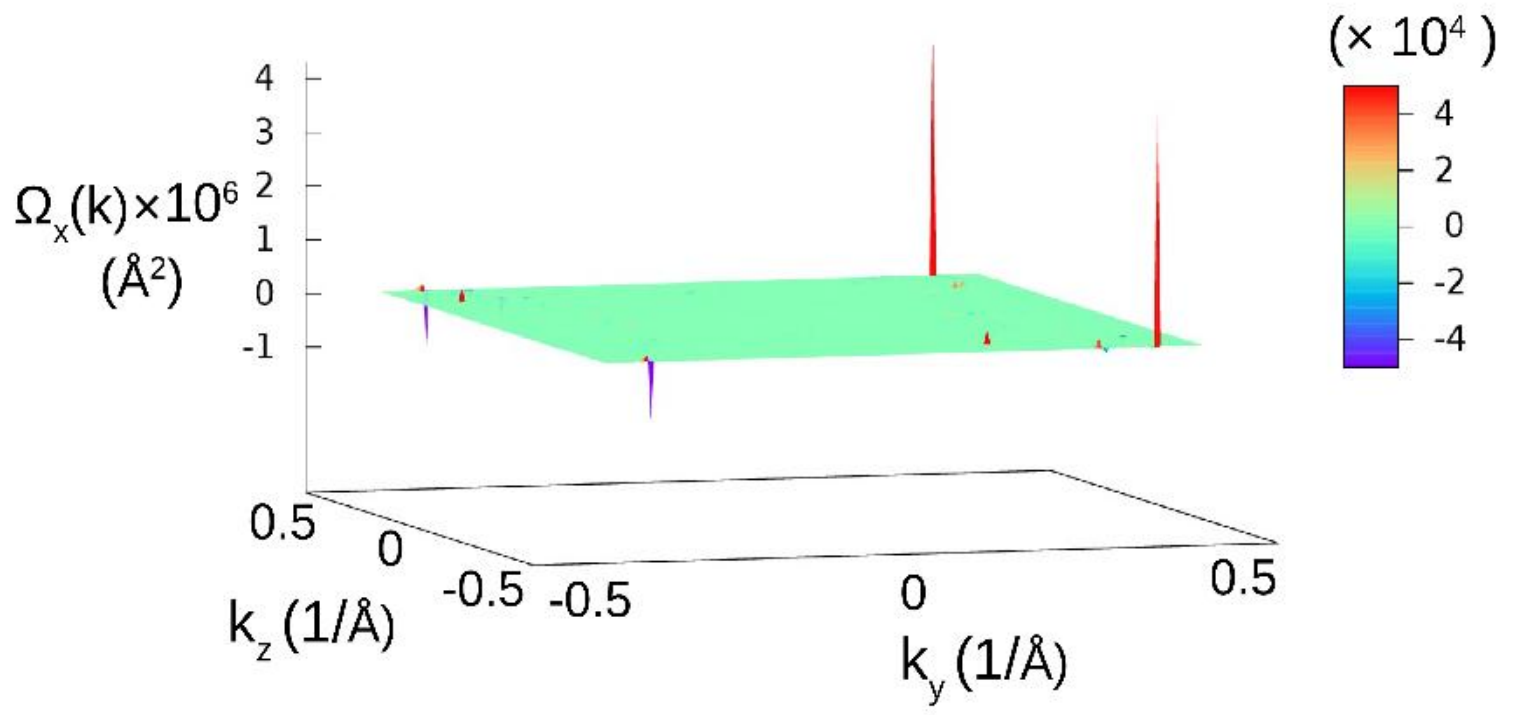

**Figure S8.** Berry curvature component along the direction of magnetic field $[110]$ for the highest occupied band with band crossing as shown in Figure S9.

As we can see clearly from Figure S9(a) and (b), where we present the band structures plotted along two different directions close to the Fermi level, the slopes of the bands (or the velocities) are more along ky (or the $[1\bar{1}0]$ in real space) direction than along kz (or [001] in real space) direction (see blue coloured bands). As per the third term in the right hand side of the equation of motion (equation (5) in main manuscript), the contribution from Berry curvature will come from its component parallel to the magnetic field. Thus, it is same for both $[1\bar{1}0]$ and $[001]$ directions. However, as the velocities of the charge carriers are more along $[1\bar{1}0]$ than along $[001]$ (e.g. the average band velocity for the four bands shown in blue in Figure S9 is $1.97 \times 10^5$ m/s along $[1\bar{1}0]$ whereas the same along $[001]$ is $0.76 \times 10^5$ m/s), the presence of LPMR along $[1\bar{1}0]$ can therefore be explained.

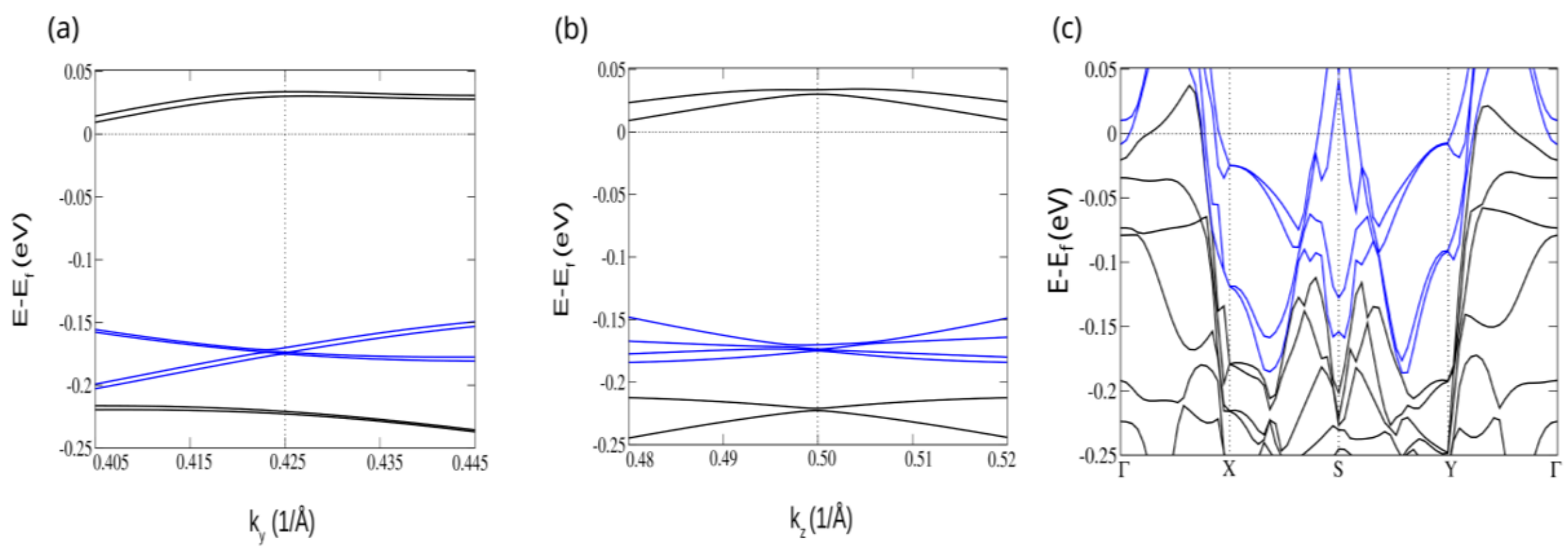

**Figure S9.** Energy dispersion around one of the k-points (shown by vertical line) which host the Berry curvature peaks along (a) $k_y$ and (b) $k_z$-direction. Fermi level is set at zero. Blue coloured bands are considered for band velocity calculations (see text). In (c) the same blue bands are shown crossing the Fermi level in high symmetry directions.

It should be noted that, even if the Weyl node (considered here, Figure S9) is located below the Fermi level, the bands constituting the node remain in close proximity in energy as they cross the Fermi level (see Figure S9(c)), giving rise to significant Berry curvature peaks at the Fermi surface. Thus, the observed Berry-curvature hotspots at EF are remnants of the underlying band topology, even if the Weyl nodes themselves do not lie exactly at the Fermi level. The anisotropy in Fermi velocities is also evident from the anisotropy in Fermi surfaces perpendicular to the $[1\bar{1}0]$ and [001] directions.

## 7. Berry curvature induced positive MR

In order to study the electrical transport properties in real materials, we should integrate the velocity of all electrons in the $\boldsymbol{k}$-space to get the current density. The current density expression with modified density of states is given by eq. 8.[14]

$$\boldsymbol{j} = -e\int \frac{d\boldsymbol{k}}{(2\pi)^3}\left[[\boldsymbol{v}_k f_0 + \frac{e}{D(\boldsymbol{k})}\tau(\boldsymbol{k})\frac{\partial f_0}{\partial \varepsilon}(\boldsymbol{E}.\boldsymbol{v}_k)\boldsymbol{v}_k + \frac{e}{\hbar}\boldsymbol{E}\times\boldsymbol{\Omega} f_0\right], \tag{8}$$

where $e$ is the elementary charge, $\boldsymbol{k}$ is the crystal momentum, $\int \frac{d\boldsymbol{k}}{(2\pi)^3}$ is integration over the Brillouin zone, $f_0$ represents the equilibrium Fermi-Dirac distribution function, $\boldsymbol{v_k}$ is the group velocity of Bloch electrons, $\tau(\boldsymbol{k})$ is the relaxation time, $\boldsymbol{E}$ is applied electric field, $\boldsymbol{\Omega}$ is the Berry curvature in momentum space, and $D(\boldsymbol{k}) = 1 + \frac{e}{\hbar}\boldsymbol{B}.\boldsymbol{\Omega}(\boldsymbol{k})$, is the berry-curvature induced phase space correction factor which modifies the density of states in the presence of magnetic field $B$.

The first term is the current in equilibrium state, which should be zero. The second term is the current induced by the electrical field. The third term is the intrinsic anomalous Hall current arising from Berry curvature. Since,

$$\boldsymbol{j}_i = \sigma_{ij}E_j, \tag{9}$$

setting $\boldsymbol{E} = E_x\hat{x}$ gives $\boldsymbol{j}_x = \sigma_{xx}E_x$ and $\boldsymbol{j}_y = \sigma_{yx}E_x$. Thus, from the second term of Eq. 8, we get the following expression for longitudinal and ordinary Hall conductivities given by eqs. 10 and 11.

$$\sigma_{xx} = -e^2\int \frac{d\boldsymbol{k}}{(2\pi)^3}\tau(\boldsymbol{k})\frac{\partial f_0}{\partial \varepsilon}\frac{1}{D(\boldsymbol{k})}v_x^2(\boldsymbol{k}), \tag{10}$$

$$\sigma_{yx}^O = -e^2\int \frac{d\boldsymbol{k}}{(2\pi)^3}\tau(\boldsymbol{k})\frac{\partial f_0}{\partial \varepsilon}\frac{1}{D(\boldsymbol{k})}v_x(\boldsymbol{k})v_y(\boldsymbol{k}), \tag{11}$$

Anomalous Hall conductivity can be derived using the third term in eq. 8.

$$\sigma_{yx}^{A} = \frac{e^2}{\hbar}\int \frac{d\boldsymbol{k}}{(2\pi)^3}\boldsymbol{\Omega}_z f_0, \tag{12}$$

where $\boldsymbol{\Omega}_z$ is the $z$-component of the Berry curvature in momentum space. The Berry curvature enters the anomalous velocity $\boldsymbol{v} = \frac{e}{\hbar}\boldsymbol{E} \times \boldsymbol{\Omega}$, such that for an electric field applied along $x$, the transverse Hall current along $y$ is governed by $\boldsymbol{\Omega}_z$. Thus, only the Berry curvature component perpendicular to the transport plane contributes to the anomalous Hall conductivity.

Expanding $\frac{1}{D(\boldsymbol{k})} = 1 - \frac{e}{\hbar}\boldsymbol{B}\boldsymbol{\Omega}_z(\boldsymbol{k})$, eq. 10 becomes:

$$\sigma_{ii}(B) \approx \int v_i^2(\boldsymbol{k}) - \frac{e}{\hbar}B\int v_i^2(\boldsymbol{k})\boldsymbol{\Omega}_z(\boldsymbol{k}), \tag{13}$$

So, MR depends on $\int v_i^2(\boldsymbol{k})\boldsymbol{\Omega}_z(\boldsymbol{k})$. Therefore, in an anisotropic system, direction-dependent band velocities lead to different effective contributions to MR, even when the Berry-curvature contribution inferred from the Hall response remains isotropic.

## 8. Direction-dependent Hall resistance measurements

Figure S10 schematically illustrates the geometry used for direction-dependent Hall in-plane resistance measurements. For **I** || $[\mathbf{001}]$, current is applied between contacts 2-4 and Hall voltage is measured between 1-3 $(\mathbf{R_{xy}} = \mathbf{V_{13}/I_{24}})$. For **I** || $[\mathbf{1\bar{1}0}]$, current is applied between contacts 1-3, and Hall voltage is measured between 2-4 $(\mathbf{R_{xy}} = \mathbf{V_{24}/I_{13}})$.

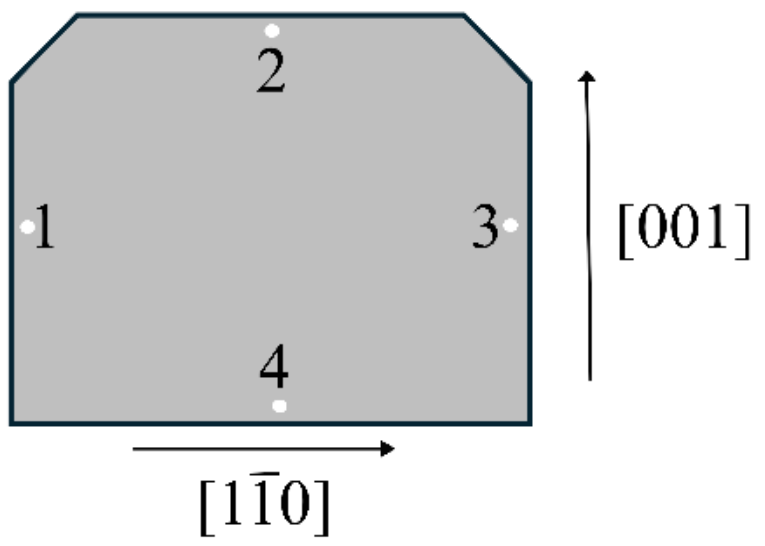

**Figure S10.** Top-view schematic of the sample showing contact configuration and crystallographic directions. For $I$ || [001], current is applied between contacts 2-4 and Hall voltage is measured between 1-3 $(R_{xy} = V_{13}/I_{24})$. For $I$ || $[1\bar{1}0]$, current is applied between contacts 1-3 and Hall voltage is measured between 2-4 $(R_{xy} = V_{24}/I_{13})$.

Figure S11(a) and (b) show the direction-dependent $R_{XY}$ at several temperatures. Figure S11(c) shows the ordinary Hall coefficient $R_0(T)$, and Figure S11(d) shows the maximum anomalous Hall resistance contribution $R_S M(T)$. In contrast to direction-dependent MR data, we obtained identical $R_{XY}$ vs. $H$ measured along [001] and [$1\bar{1}0$]. The extracted ordinary and anomalous Hall contributions exhibit identical temperature dependence in both directions.

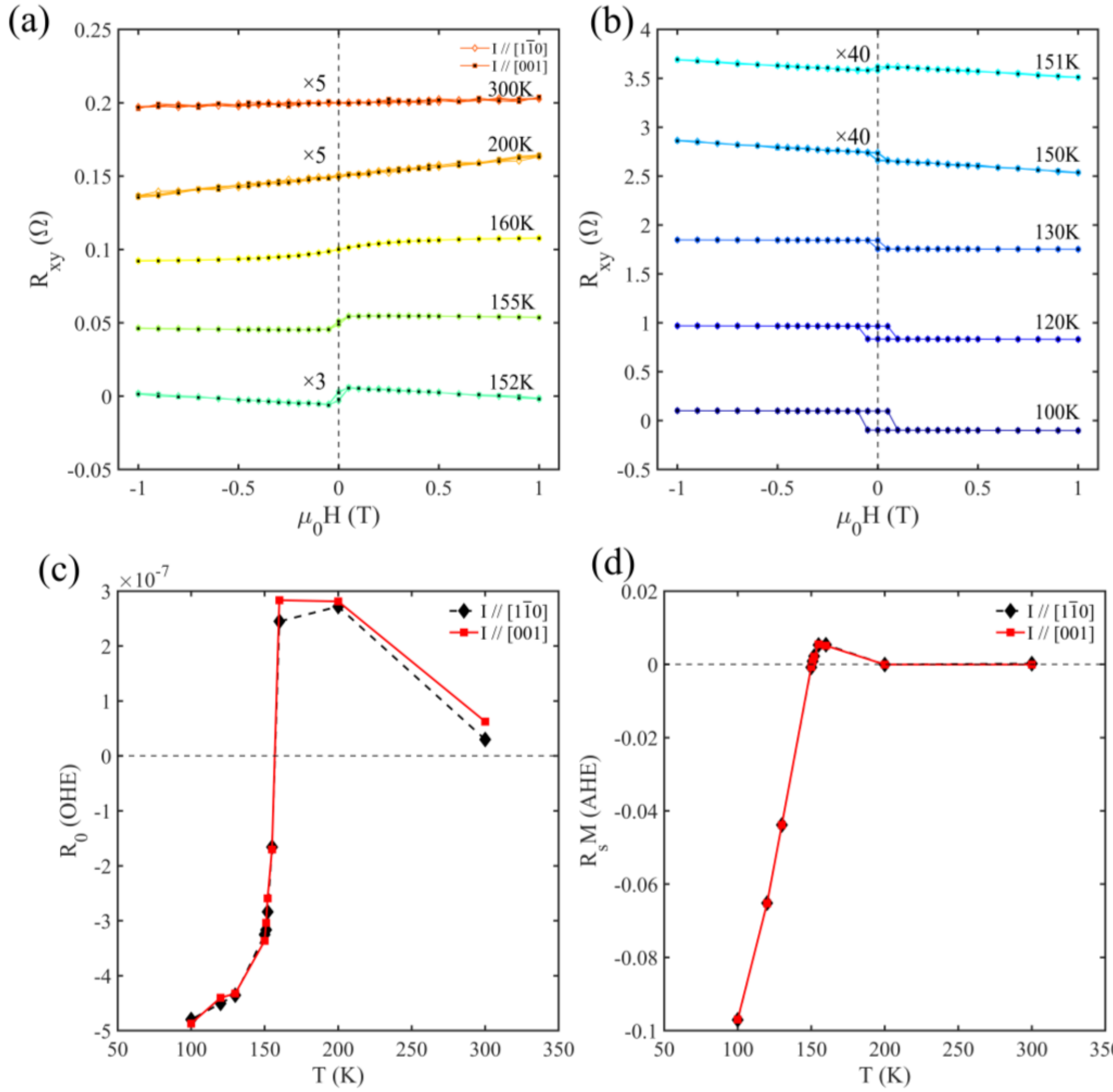

**Figure S11.** (a) and (b) Direction-dependent Hall resistance as a function of applied magnetic field at several temperatures for $I$ || [$1\bar{1}0$], and $I$ || [001]. (c) Direction-dependent Ordinary Hall coefficient, and (d) Total anomalous Hall effect part (Rs*M) as a function of temperature.

The sign reversal of both OHE and AHE as a function of temperature indicates a crossover in the dominant carrier type and a reconstruction of the Berry curvature contribution near the Fermi level. It occurs at the same characteristic temperature for both crystallographic directions. OHE changes sign from positive to negative at 155K, while AHE reverses its sign at 150K, similar to the average Hall response (Fig. 3 maintext). This directional independence of the Hall response, in contrast to the anisotropic longitudinal transport, suggests that while scattering mechanisms are strongly anisotropic, the underlying electronic structure and Berry curvature contributions remain effectively isotropic in the plane.

We can easily understand the identical Hall resistance vs. T and the magnitude from eqs. (11) and (12). In particular, ordinary Hall resistance in eq. (11) has both orthogonal charge velocity components and the Berry curvature component $\boldsymbol{D(k)}$ is the identical for $\boldsymbol{\sigma_{xy}}$ and $\boldsymbol{\sigma_{yx}}$. Consequently, the ordinary hall contribution $\boldsymbol{R_0(T)}$ remains nearly identical for both directions. On the other hand, anomalous Hall resistance in eq. (12) has the pure contribution from the Berry curvature component along the magnetic field direction (z-direction). This is orthogonal to both in-plane directions. Thus, it exactly has the same contribution in both in-plane directions.

## 9. MR measurement on 25 nm SRO film on DSO

Figure S12 shows the directional MR measurement on a thinner, 25 nm SRO/DSO film. This film reproduces the same directional MR sign reversal along $[1\bar{1}0]$ observed in the primary 50 nm film. This demonstrates the reproducibility of the effect across different film thicknesses and argues against a purely sample-specific, disorder-driven, or domain-driven origin.

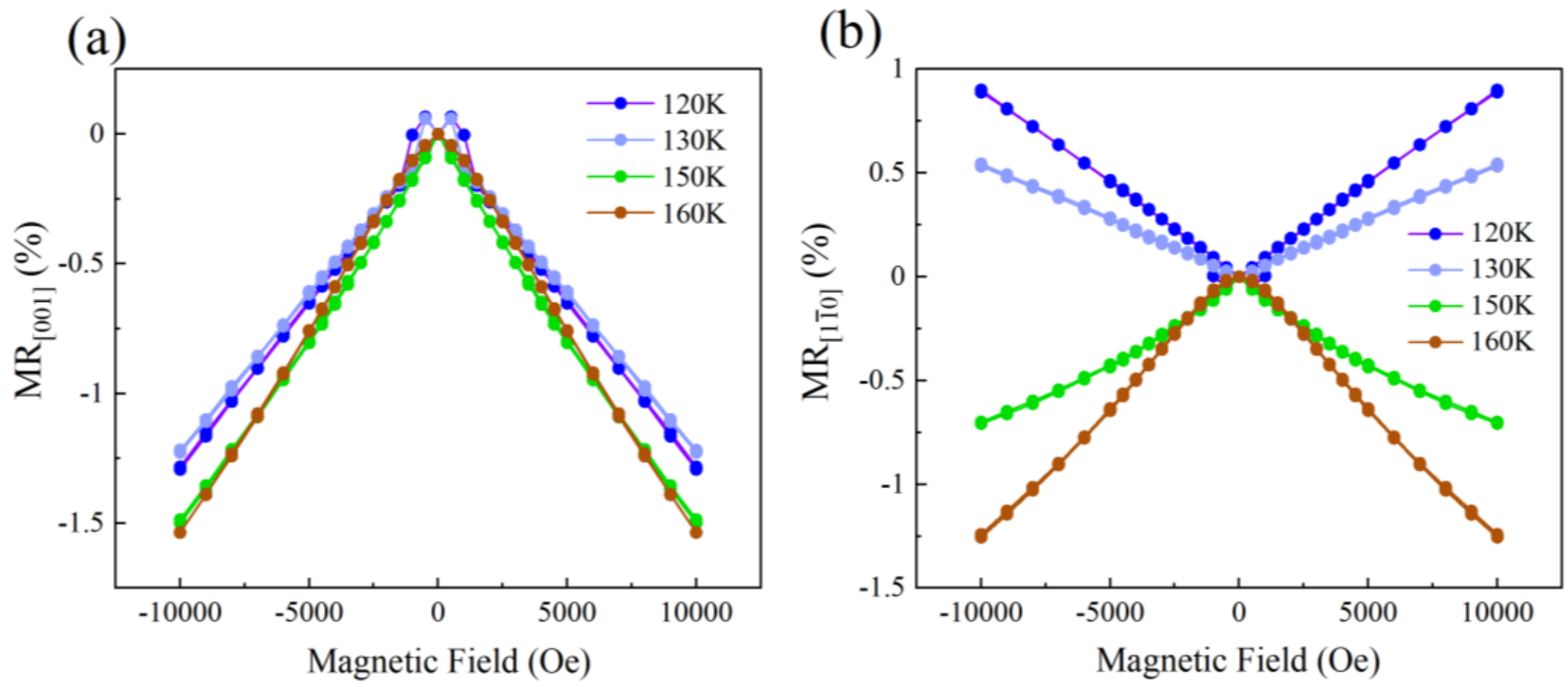

Figure S12. Magnetic field $(\mu_0 H)$ dependence of magnetoresistance (MR, symbols) along two in-plane directions (a) $[001]$, and (b) $[1\bar{1}0]$ of a 25nm SRO-DSO film.

## 10. Verification of directional MR measurement on 50 nm SRO film on DSO using linear four-probe geometry

Figure S13 shows the directional MR measurement on our primary 50 nm SRO/DSO film using linear four-probe contacts along the $[001]$ and $[1\bar{1}0]$ directions at selected temperatures (120 K, 130 K, and 150 K). The four-probe measurement reproduces the similar directional MR anisotropy observed in the Van der Pauw geometry, confirming that the anisotropy is intrinsic to the film and not an artifact of the Van der Pauw measurement geometry.

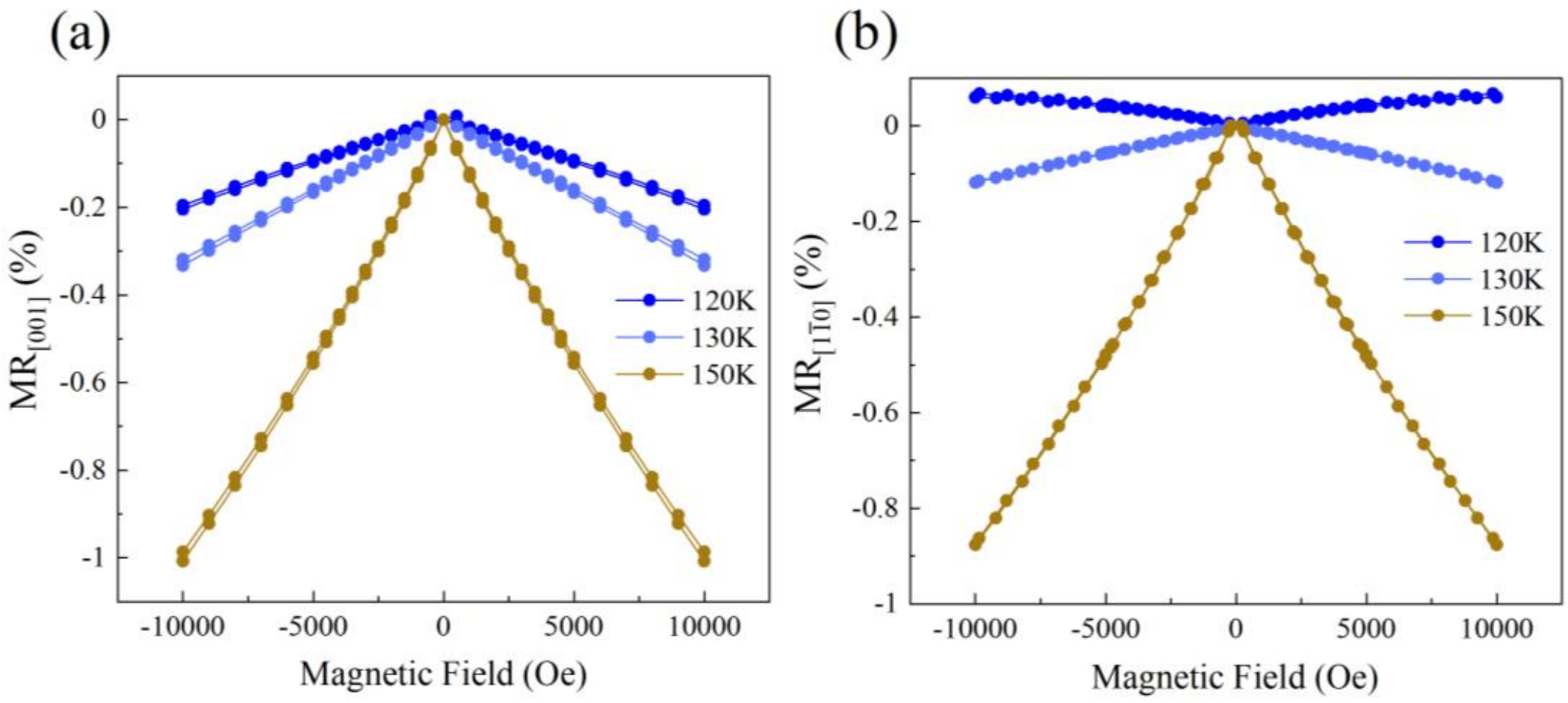

Figure S13. Magnetic field $(\mu_0 H)$ dependence of magnetoresistance (MR) on 50 nm SRO film on DSO along two in-plane directions (a) $[001]$, and (b) $[1\bar{1}0]$ using linear four-probe contacts.